\documentclass[twocolumn]{aastex62}
\pdfoutput=1
\usepackage{amsmath,amsfonts,amssymb,mathrsfs,graphicx,color}
\usepackage[squaren]{SIunits}
\usepackage{verbatim}
\usepackage{enumerate}
\usepackage{graphicx}
\usepackage{xcolor}
\usepackage{slashed}
\usepackage[utf8]{inputenc}
\usepackage[percent]{overpic}
\usepackage{natbib}
\usepackage{multirow}
\usepackage{float}
\usepackage{booktabs}
\usepackage[titletoc,title]{appendix}
\usepackage{graphicx}
\graphicspath{{figures/}}

\newcommand{\ie}{{\it i.e.}}
\newcommand{\eg}{{\it e.g.}}

\newcommand{\cf}{{\it cf.}}

\newcommand{\eq}{Eq.}

\newcommand{\Fig}{Fig.}

\newcommand{\Sec}{Section}
\newcommand{\Secs}{Sections}
\newcommand{\App}{Appendix}

\newcommand{\Tab}{Tab.}

\newcommand{\equ}[1]{\eq~(\ref{eq:#1})}
\newcommand{\figu}[1]{\Fig~\ref{fig:#1}}
\newcommand{\tabl}[1]{\Tab~\ref{tab:#1}}
\newcommand{\sect}[1]{\Sec~\ref{sec:#1}}
\newcommand{\append}[1]{\App~\ref{sec:#1}}

\begin{document}

\title{Interpretation of the diffuse astrophysical neutrino flux in terms of the blazar sequence}

\author{Andrea Palladino}
\email{andrea.palladino@desy.de}
\affiliation{DESY, Platanenallee 6, 15738 Zeuthen, Germany}

\author{Xavier Rodrigues}
\email{xavier.rodrigues@desy.de}
\affiliation{DESY, Platanenallee 6, 15738 Zeuthen, Germany}

\author{Shan Gao}
\email{shan.gao@desy.de}
\affiliation{DESY, Platanenallee 6, 15738 Zeuthen, Germany}

\author{Walter Winter}
\email{walter.winter@desy.de}
\affiliation{DESY, Platanenallee 6, 15738 Zeuthen, Germany}

\date{\today}


\begin{abstract}   
We study if the diffuse astrophysical neutrino flux can come from blazar jets -- a subclass of Active Galactic Nuclei (AGNs) -- while it, at the same time, respects the blazar stacking limit based on source catalogs and is consistent with the observation from TXS 0506+056.
We compute the neutrino flux from resolved and unresolved sources using an averaged, empirical relationship between electromagnetic spectrum and luminosity, known as the {\em blazar sequence}, for two populations of blazars (BL Lacs and FSRQs). Using a source model with realistic neutrino flux computations, we demonstrate that blazars can indeed power the diffuse neutrino flux at the highest energies and obey the stacking limit at the same time, and we derive the conditions for the baryonic loading (proton versus $\gamma$-ray luminosity) evolving over the blazar sequence. Under the hypothesis that low-luminosity blazars power the diffuse astrophysical neutrino flux, we find that the dominant contribution of the diffuse flux up to PeV energies must come from unresolved BL Lacs with baryonic loadings larger than about $10^5$ -- while only a very small contribution may come from resolved high-luminosity BL Lacs or FSRQs, which can be  directly tested by the stacking limit. We find that the blazar TXS 0506+056 is on the verge of these populations in our baseline scenario, at a relatively high luminosity and redshift; as a consequence we predict about 0.3 $\gamma$-ray-neutrino associations per year from the whole population, dominated by BL Lacs with $L_\gamma \simeq 10^{45} \, \mathrm{erg/s}$ and $z \sim 0.1$.
\end{abstract}

\section{Introduction}
\label{sec:introduction}

The discovery of a diffuse flux of high-energy neutrinos by IceCube~\citep{icescience} has triggered many questions, one of the most relevant being \lq\lq what is the source of the high-energy neutrinos detected by IceCube?\rq\rq. Several candidate classes have been proposed to predict neutrino fluxes and possibly interpret the IceCube data, such as blazars ~\citep{bl1,bl2,bl3,bl4,bl5,Righi:2018xjr,Righi:2018hhu,Murase:2018iyl}, Gamma-Ray Bursts (GRBs)~\citep{grb1,Waxman:1997ti,grb2,grb3,grb4,grb5}, starburst galaxies~\citep{sb1,sb2,sb3}, the cores of Active Galactic Nuclei (AGNs,~\citet{Stecker:2013fxa}), and dark matter decay~\citep{dm1,dm2}. Some of them are already strongly constrained by measurements, based on the lack of correlations between high-energy neutrinos and known sources. For example, it is known that the contribution of GRBs to the IceCube neutrino signal is of the order of a few percent~\citep{Aartsen:2017wea}, whereas the contribution of starburst galaxies may be larger but is also insufficient to explain the observed neutrinos, due to constraints from the extragalactic $\gamma$-ray background~\citep{Murase:2013rfa,stacksbg}. A certain contribution to the neutrino flux may also come from the Galactic plane \citep{pal1,pal2} but the latest measurements of IceCube and ANTARES suggest that this contribution cannot be greater than $\sim 10\%$ \citep{Albert:2018vxw}.

In this work, we study the contribution of blazars to the astrophysical neutrino flux. Blazars are a sub-class of AGNs, which are supermassive black holes that accrete matter from the host galaxy, launching relativistic jets and emitting non-thermal radiation. In the case of blazars, the relativistic jet points in the direction of Earth. Assuming that cosmic rays (CRs) may be accelerated in the jet, we calculate the neutrino emission from CR interactions with the blazar photon fields. As the target particles for CRs, the distributions of those photons in the blazar family are represented by a series of spectra called \textit{blazar sequence}~\citep{Ghisellini:2017ico}, based on an empirical relation between the luminosity of a blazar and the photon spectrum emitted.


Considering the high-energy starting events (HESE) in IceCube, the contribution of resolved blazars to the astrophysical neutrino flux cannot be larger than $\sim 20\%-25\%$~\citep{stackbl} based on the missing association with sources in $\gamma$-ray catalogs, or even less based on theoretical considerations \citep{Palladino:2017aew}. On the other hand, there have been indications of associations of individual neutrino events with AGNs~\citep{Resconi:2016ggj,Padovani:2016wwn,Kadler:2016ygj}, which means that it is plausible that a few of IceCube's events stem from resolved blazars. More recently, direct evidence has been found of a correlation between IceCube neutrinos and the object TXS 0506+056 \citep{IceCube:2018dnn,IceCube:2018cha}, which represents a breakthrough discovery in multi-messenger astrophysics. This observation may support  the AGN origin of the diffuse neutrino flux, possibly powered by unresolved objects.

In this work we show that low-luminosity blazars can provide the dominant contribution to the high-energy neutrinos with energy between a few hundreds of TeV and a few PeV, while the contribution of very bright sources to the neutrino flux must be highly suppressed in order to respect the blazar stacking limit~\citep{stackbl}. This result allows us to draw conclusions about the possible baryonic loading of blazars evolving over the blazar sequence.
We also discuss explicitly the role of TXS 0506+056, and what we can learn from the population study for future such observations.

In \sect{neutrino_production} we evaluate the neutrino spectra from blazars numerically, taking into account the differences between the two blazar sub-classes, namely BL Lacs and Flat-Spectrum Radio Quasars (FSRQs). In \Sec~\ref{sec:sourcedist} we discuss the cosmic evolution of BL Lacs~\citep{Ajello:2013lka} and FSRQs~\citep{Ajello:2011zi}, which is necessary to compute the expected diffuse neutrino flux at Earth. We present our results in \sect{res}, where we compute the diffuse flux of neutrinos under three different hypotheses: \textit{i)} constant baryonic loading (defined as the power ratio of injected non-thermal protons to the emitted gamma-rays; see \equ{xidef}.), \textit{ii)} proportionality between the luminosity of blazars in neutrinos ($L_\nu$) and in $\gamma$-rays ($L_\gamma$), and \textit{iii)} a baryonic loading that scales with the source luminosity. A discussion is presented in \sect{discussion},  focusing on \textit{i)} source characteristics predicted by the model as a function of $\gamma$-ray luminosity, \textit{ii)} implications of the multiplet constraint, \ie, the non-observation of two neutrino events from the same source, \textit{iii)} implications of our model to potential neutrino sources such as TXS 0506+056 and \textit{iv)} PKS 0502+049 \citep{He:2018snd}. 
We conclude the work with \sect{conclusion}.


\section{Methods}
\label{sec:methods}

We now describe the methods used to compute the neutrino flux from blazars. In \sect{neutrino_production} we present the radiation model for the calculation of neutrino spectra across the blazar sequence, in \sect{sourcedist} we address the cosmic evolution of blazars, in \sect{neutrinoflux} we explain the procedure to calculate the diffuse neutrino flux observed at Earth, and in \sect{txsintro} we introduce the special blazar TXS 0506+056.

\subsection{Neutrinos from the blazar sequence}
\label{sec:neutrino_production}

\begin{figure}[tbp!]
\centering
\includegraphics[width=0.85\linewidth]{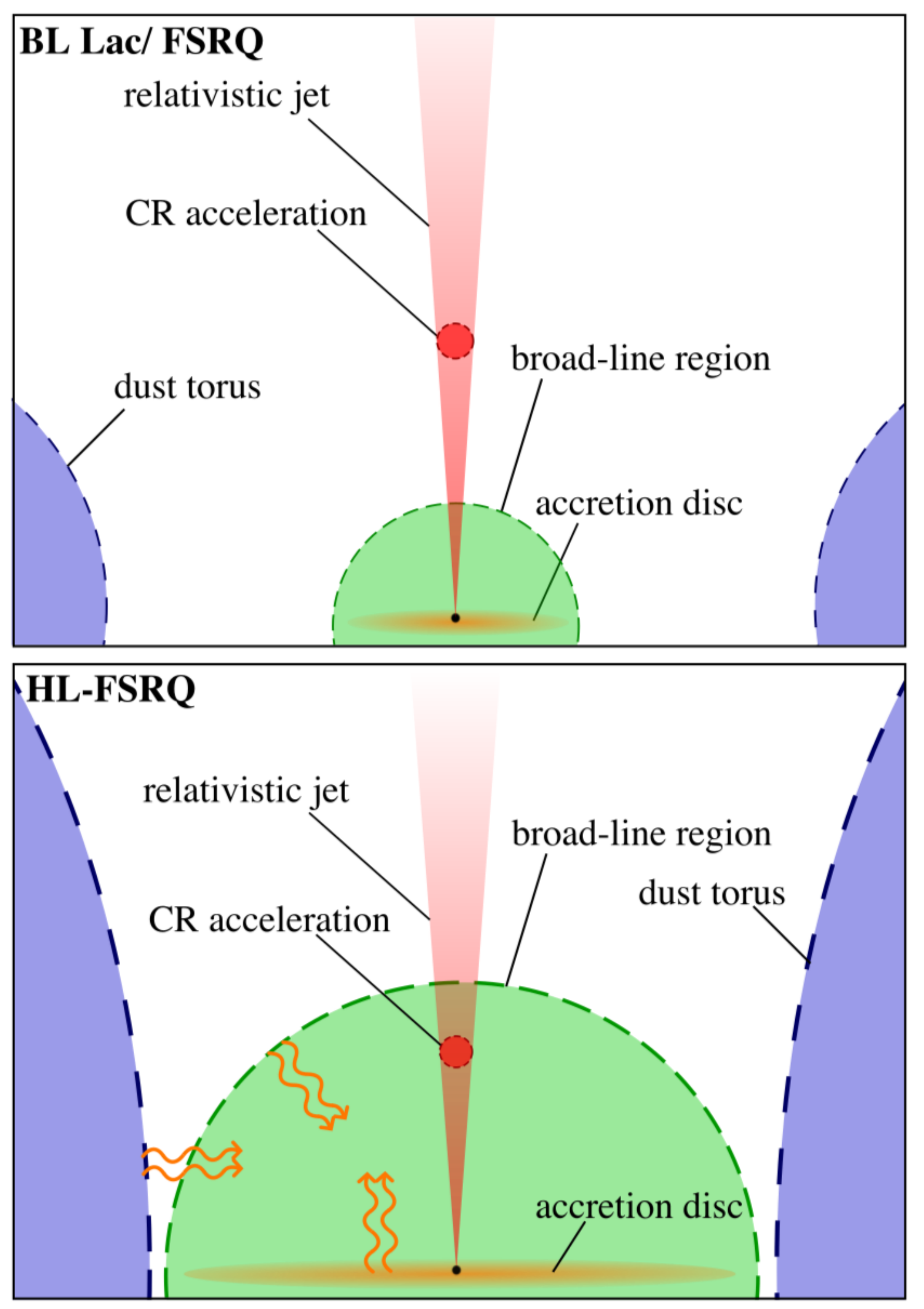}
\caption{Schematics of the blazar model (not to scale). Top: BL Lac or low-luminosity FSRQ. In these sources the relativistic jet blob, where CRs are injected, lies outside the broad line region (BLR), and therefore CRs hardly interact with the photon fields produced by the accretion disk or the gas surrounding the BLR -- which are de-boosted in the blob frame. Bottom: high-luminosity FSRQ (HL-FSRQ), where the acceleration region lies in the BLR and the external photon spectra are relevant for the interactions in the blob frame. If we assume that the luminosity of the source increases with the size of the accretion disk, then in FSRQs brighter than $L_\gamma=2.4\times10^{48}~{\rm erg/s}$ the jet blob lies inside the BLR in this model, and the efficiency of photo-hadronic interactions is increased by the external photon fields of the FSRQ. Figure adopted from~\citet{Rodrigues:2017fmu}.}
\label{fig:model}
\end{figure}

The calculation of neutrino production in blazars follows the same methods used in~\citet{Rodrigues:2017fmu}. 
Some details of the present calculation, including key differences to the previous implementation of the model, are discussed in \append{source_model}. The basic assumption is that CR protons are accelerated in the relativistic jet of the blazar during a period of flaring activity. We assume the same jet speed for all the sources, corresponding to a Lorentz factor $\Gamma=10$. We assume the source is in a persistent flaring state, corresponding to a duty cycle of 100\% (see \sect{txsintro} for a discussion on the implications of different duty cycle assumptions). Cosmic rays are injected isotropically during a period of time corresponding to a typical flare, $t'=10^{6}~{\rm s}$,\footnote{Primed quantities will refer to the rest frame of the jet plasma, while unprimed quantities may refer to the source or the observer's frame.} in a spherical region of size $r'_{\rm blob}=ct'_{\rm flare}$. This is the region represented in dark red in \figu{model}.

\begin{figure*}[t]
\includegraphics[width=\linewidth]{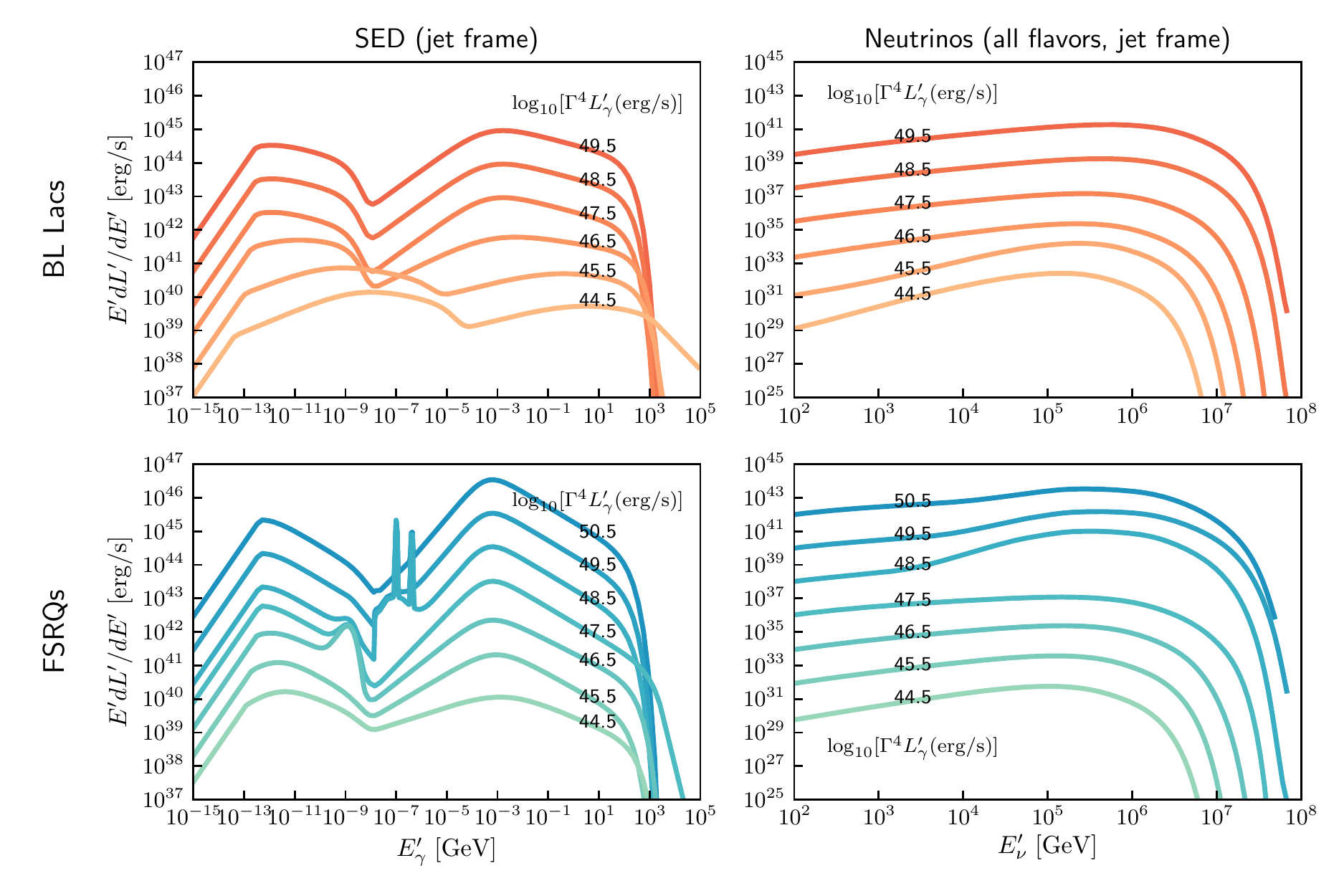}
\caption{Left: spectral energy distributions (SEDs) for BL Lacs (top) and FSRQs (bottom) used in this work, given in the rest frame of the jet. The non-thermal SEDs are those from the blazar sequence~\cite{Ghisellini:2017ico}, and the external components of the energy spectra are based on~\cite{bl3}. Right: neutrino luminosity spectra calculated for each blazar luminosity, for a baryonic loading of $\xi=1$, also in the rest frame of the jet (see \append{source_model} for details). The label of each curve indicates the logarithm of the $\gamma$-ray luminosity in the black hole frame in erg/s, given by $L_\gamma=\Gamma^4L'_\gamma$, where $\Gamma=10$ is the bulk Lorentz factor of the jet.}
\label{fig:neutrinos_sequence}
\end{figure*}

Cosmic rays are assumed to undergo second-order Fermi acceleration, with an acceleration rate given by
\begin{equation}
t'^{-1}_{\rm acc}(E',B') \,=\, \eta\frac{ZeB'c}{E'},
\label{eq:eta}
\end{equation}
where $Z$ is the atomic number of the nucleus being accelerated, $E'$ is its energy, and $B'$ is the magnetic field strength in the jet (assumed to scale with the blazar luminosity, see \append{source_model}). Moreover, $\eta$ is the acceleration efficiency, which is assumed to be $\eta=10^{-3}$ in this work.  That low value of acceleration efficiency is motivated by matching the primary energy to the cutoff of the IceCube neutrinos, and implies that there is no connection to ultra-high-energy cosmic rays (as \eg\ in~\citet{Rodrigues:2017fmu}, where a higher value of $\eta$ was used). We assume CRs are accelerated to a power-law energy distribution in the jet:
\begin{equation}
\frac{dN}{dE'} \,=\, E'^{-2} \, \exp\left({-\frac{E'}{E'_{\rm max}}}\right) \, ,
\end{equation}
where $E'_{\rm max}$ is the maximum energy achieved by the cosmic rays in the source. This is the energy at which cooling or interaction processes become more efficient than acceleration, or the acceleration becomes limited by the size of the blob.

The interaction of CRs with a target photon field is simulated using the \textsc{NeuCosmA} code~\citep{Baerwald:2011ee}. The main process responsible for neutrino emission is photo-meson production, where in most cases the neutrinos produced carry around 5\% of the energy of the primary. The amount of CRs injected in each source is a free parameter of the model; we quantify this quantity by means of the {\bf baryonic loading} of the source, defined as 
\begin{equation}
\xi \equiv \frac{L_{\rm CR}}{L_{\gamma}} \label{eq:xidef}  \, ,
\end{equation}
where $L_{\rm CR}$ is the total luminosity of injected CRs and $L_{\gamma}$ is the $\gamma$-ray luminosity of the source, defined here in the range $0.1-100~{\rm GeV}$, roughly corresponding to the Fermi-LAT observation range.

The spectral energy distribution (SED) of each source (used as the target photon field for hadronic interactions) is determined according to the most recent implementation of the blazar sequence~\citep{Ghisellini:2017ico}, an empirical relation based on the Fermi 3LAC blazar catalog~\citep{Ackermann:2015yfk} that attributes  an average SED to each blazar luminosity $L_\gamma$, distinguishing between BL Lacs and FSRQs. The SEDs are shown in the left panels of \figu{neutrinos_sequence}: the upper panels refer to BL Lacs, and the bottom panels to FSRQs. All SEDs with $L_\gamma>10^{48}~{\rm erg/s}$ and $L_\gamma<10^{44}~{\rm erg/s}$ have been extrapolated by renormalizing the brightest and dimmest SEDs (respectively) provided in~\citet{Ghisellini:2017ico}.  As we can see in the top left panel of \figu{neutrinos_sequence}, high-frequency peaked BL Lacs (HBLs), \ie, BL Lacs with synchrotron and inverse Compton peaks at higher energies, are generally dimmer sources, while low-frequency peaked BL Lacs (LBLs) are brighter. For FSRQs, on the other hand, there is no strong relationship between the frequency of non-thermal emission bands and the jet luminosity (bottom left panel of \figu{neutrinos_sequence}). Besides synchrotron and inverse Compton emission, FSRQs typically exhibit bright broad lines from atomic emission of the gas surrounding the accretion disk, as well as thermal bumps from an accretion disk and a dusty torus. If we assume that the size of the broad line region (BLR, green region in \figu{model}) increases with luminosity, then the radiation zone in the jet is contained within reach of the external photon fields for high-luminosity (HL) FSRQs  (bottom panel of \figu{model}, \cf, \citet{bl3}). These external fields, seen in the SEDs of the brightest FSRQs in \figu{neutrinos_sequence}, enhance CR interactions both inside the jet and during the propagation of escaping CRs through the BLR and near the dusty torus. This is taken into account by means of two additional radiation regions in the modeling of high-luminosity FSRQs (see \append{source_model}).

In the right panels of \figu{neutrinos_sequence} we show the emitted neutrino spectra obtained with our blazar model. The integral of the neutrino spectrum of each blazar yields the total neutrino luminosity of the source, $L_\nu$. This quantity is directly proportional to the baryonic loading $\xi$ of the source, since the emitted neutrinos originate from CR interactions (\figu{neutrinos_sequence} refers to the special case $\xi=1$). In general, the emitted neutrino luminosity $L_\nu$ of a blazar may be written as
\begin{equation}
L_\nu = \mathcal{K} \, L_\gamma,
\label{eq:neutrino_luminosity}
\end{equation}
where $\mathcal{K} \equiv \epsilon_\nu \times \xi$ is the product of the neutrino production efficiency and the baryonic loading, defined in \equ{xidef}. The neutrino production efficiency $\epsilon_\nu \equiv L_\nu/L_{\rm CR}$  quantifies the amount of energy from CRs converted into neutrinos in the source (which is independent of the baryonic loading). The quantity $\mathcal{K}=L_\nu/L_\gamma$ will be used throughout this work to express the ratio between the neutrino and $\gamma$-ray luminosity of a given source.

\begin{figure}[tp]
\includegraphics[width=\linewidth]{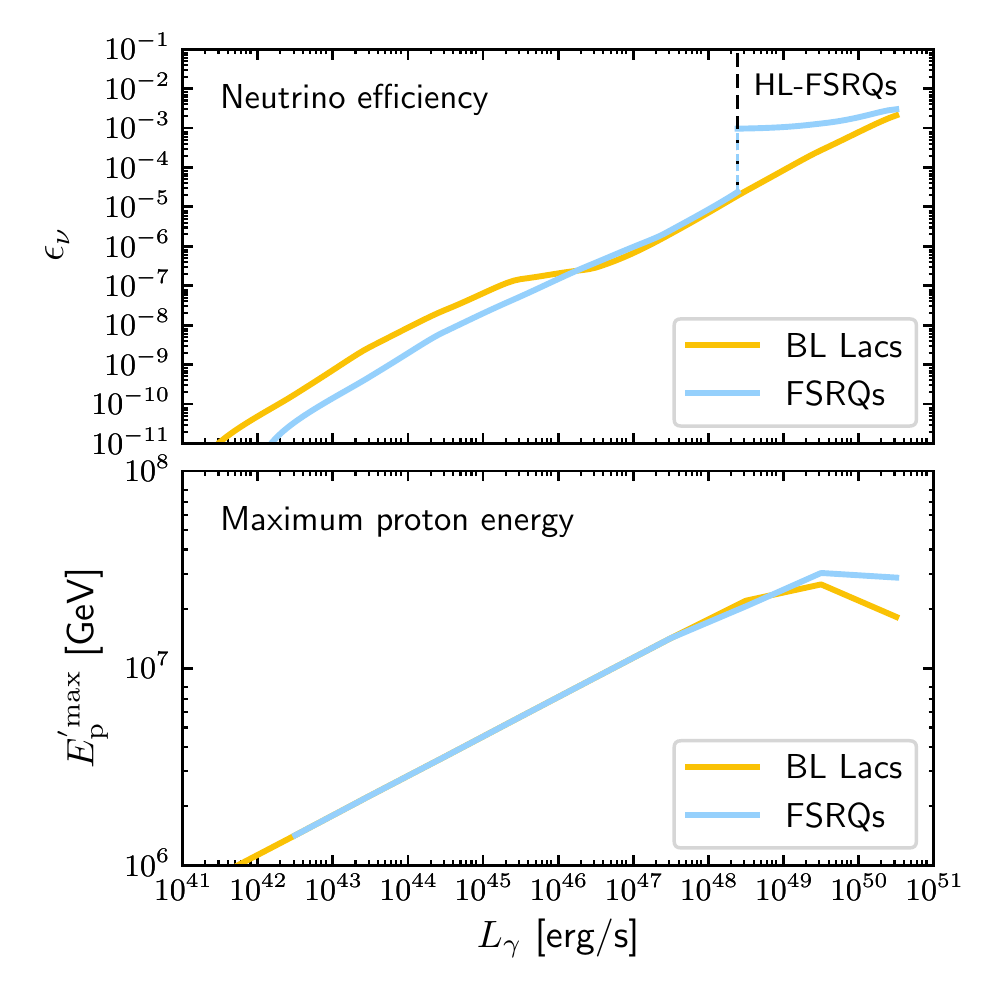}
\caption{Top: neutrino production efficiency of BL Lacs (yellow) and FSRQs (blue) of the blazar sequence as a function of the $\gamma$-ray luminosity of the source. Bottom: maximum energy of accelerated protons in the jet of BL Lacs (yellow) and FSRQs (blue) as a function of luminosity. We assume that the radiation zone in the jet is exposed to external photon fields from molecular and thermal emission in the case of FSRQs with luminosity above $3\times10^{48}~{\rm erg/s}$ (HL-FSRQs); this is responsible for the jump in neutrino efficiency (see \append{source_model}).}
\label{fig:efficiency}
\end{figure}

Using the results of \figu{neutrinos_sequence}, we can calculate the neutrino production efficiency of each source considered in this work. This is shown in the top panel of \figu{efficiency} as a function of the $\gamma$-ray luminosity of the source. The neutrino production efficiency increases monotonically with the $\gamma$-ray luminosity of the blazar, since higher luminosities imply higher photon densities in the radiation zone in the jet. Note the abrupt increase in the efficiency of FSRQs with $L_\gamma\gtrsim3\times10^{48}~{\rm erg/s}$, which is due to interactions with external fields, producing additional neutrinos.  

In the bottom panel of \figu{efficiency} we show the maximum energy achieved by protons accelerated in the jet, which is highest for blazars of $L_\gamma\sim3\times10^{49}~{\rm erg/s}$. Above this luminosity, photo-hadronic interactions of high-energy CRs dominate over acceleration, and the maximum CR energy starts decreasing with luminosity.

\begin{figure*}[t]
\centering
\includegraphics[width=0.45\textwidth,angle=0]{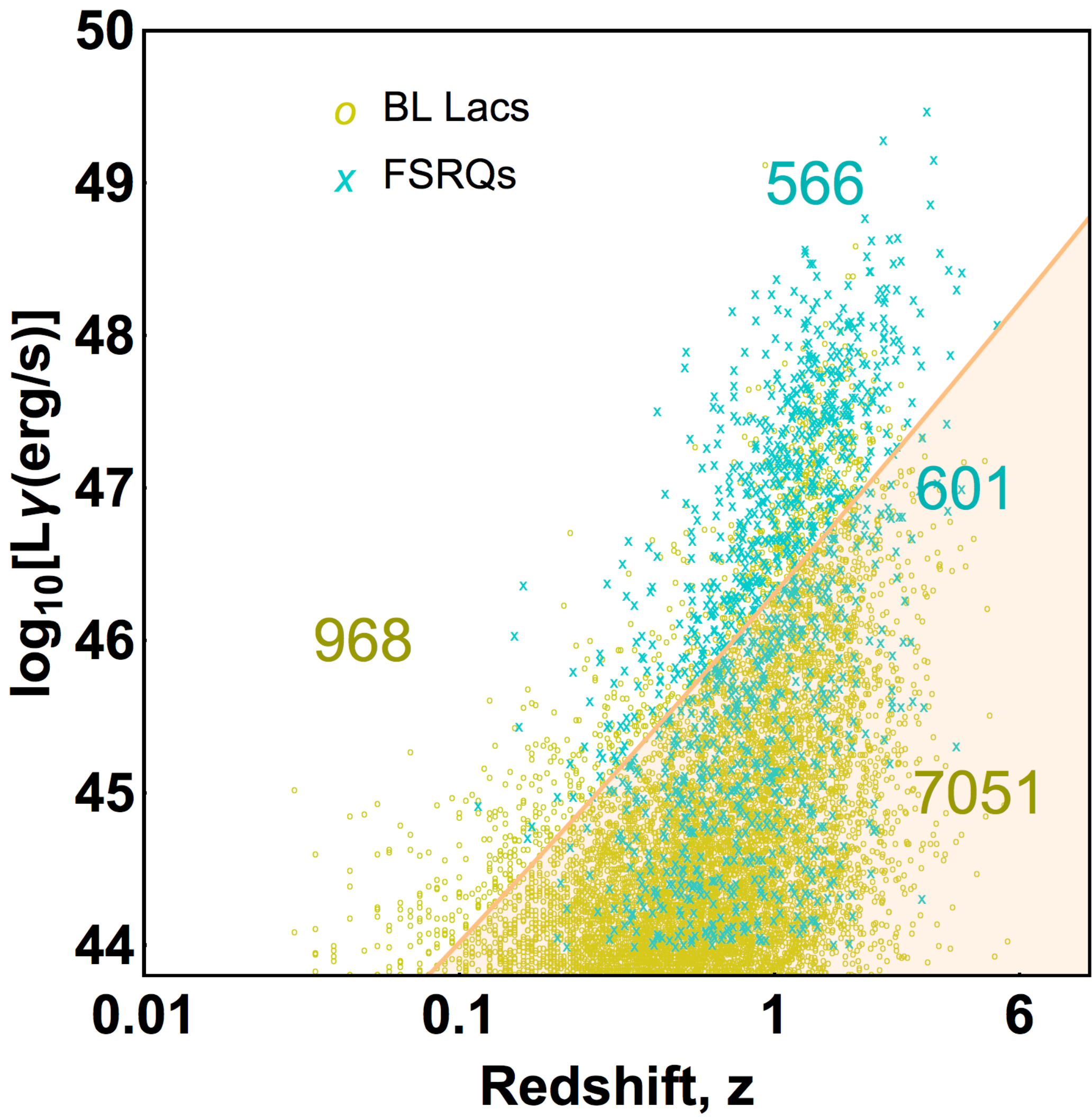}
\includegraphics[width=0.46\textwidth,angle=0]{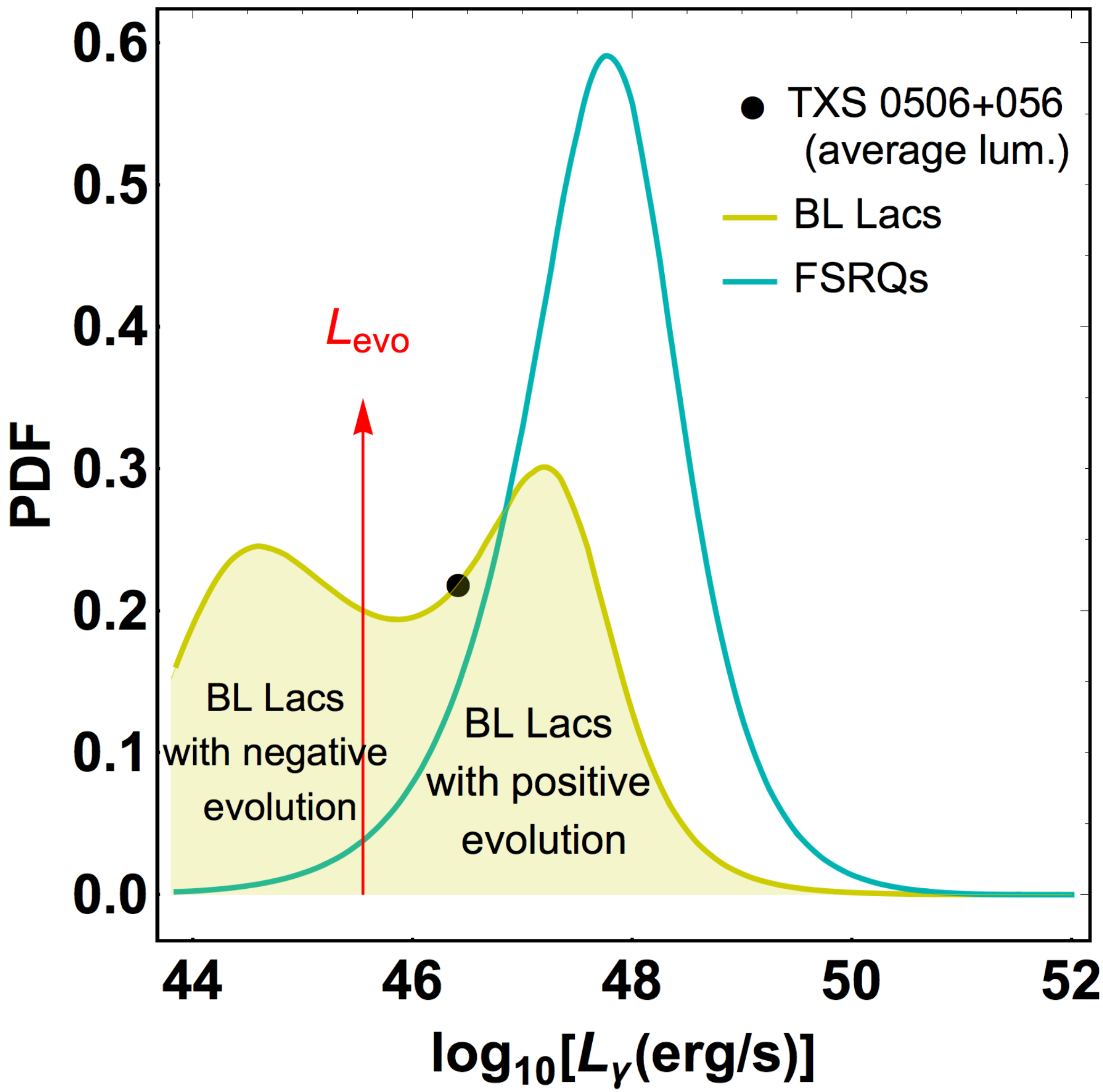}
\caption{Left panel: the cosmic evolution of BL Lacs and FSRQs as a function of redshift $z$ and logarithm of the $\gamma$-ray luminosity $L_\gamma$, following the distributions given in \citet{Ajello:2013lka,Ajello:2011zi}. The orange curve roughly separates resolved sources (above) from unresolved sources (below). The yellow and cyan numbers denote the number of resolved and unresolved BL Lacs and FSRQs respectively. Right: distribution of BL Lacs and FSRQs as a function of luminosity, obtained by integrating over the redshift. We notice that FSRQs are accumulated at high luminosity, whereas BL Lacs are characterized by two different populations. The luminosity $L_\gamma^{\text{evo}}=3.5 \times 10^{45} \text{erg/s}$ is indicated.}
\label{fig:distro}
\end{figure*}

\subsection{Source evolution}
\label{sec:sourcedist}

The calculation of the cumulative neutrino flux from blazars, reported in this work, is based on the distribution of BL Lacs and FSRQs in~\citet{Ajello:2013lka, Ajello:2011zi}. In these two works the distribution of blazars is parametrized in order to reproduce the sources observed by Fermi-LAT and the observed diffuse $\gamma$-ray background~\citep{fermi2}. The parametrization is also useful to evaluate the contribution of unresolved sources, \ie, the sources below the Fermi-LAT sensitivity threshold (but expected from a theoretical point of view) that contribute to the diffuse $\gamma$-ray flux.  

In \append{ajello} we give the details of the parameterization by Ajello et al., which we use in \sect{neutrinoflux} to compute the cumulative neutrino flux from blazars. The distribution describes the number of sources per redshift and luminosity and is characterized by 12 parameters (which are different for BL Lacs and FSRQs), reported in \tabl{distro}.\footnote{In \citet{Ajello:2013lka, Ajello:2011zi} \eq~(\ref{eqwrong}) is reported with the wrong signs of $p_1$ and $p_2$. This produces an incorrect distribution which is particularly relevant for low-luminosity BL Lacs. In this work we use this function considering the correct signs. Please note that the erratum has been reported in a later work in 2015. See the second footnote of \citet{Ajello:2015mfa}.}.

The distributions of BL Lacs and FSRQs are reported in \figu{distro}. In these figures we represent a set of 9186 sources, obtained from a Monte Carlo extraction in agreement with the distribution presented in \append{ajello}.
The total number of sources is already contained in the  parametrization by Ajello et al. We represent the source evolution as a function of redshift $z$  and luminosity $\ell \equiv \log_{10} [L_\gamma~\rm{(erg/s)}]$. The threshold separating the resolved and unresolved sources is represented by a single value of $\phi_\gamma=4 \times 10^{-12}~{\rm erg~cm^{-2}~s^{-1}}$ for simplicity, in order to reproduce the $\sim$ 1500 blazars listed in the 3LAC catalog. In reality this value depends strongly on the gamma-ray spectral index, the exposure time and the position of the source \citep{Ackermann:2015yfk}.
It is interesting to notice that about 50\% of FSRQs are resolved, whereas more than 85\% of the expected BL Lacs are not resolved but they are expected to contribute to the total flux of $\gamma$-rays and neutrinos. 

In the right panel of \figu{distro} we show the distribution of BL Lacs and FSRQs as a function of their luminosity, obtained integrating on the redshift $z$. We notice that most FSRQs have a luminosity  between $L_\gamma= 10^{46}$ and $10^{50}~\text{erg/s}$, whereas BL Lacs are characterized by two populations with different characteristics: a low-luminosity population (roughly between $L_\gamma=10^{44} - 10^{46}~\text{ erg/s}$) and a high-luminosity population above $L_\gamma \simeq 10^{46}~\text{ erg/s}$.  We note that the choice of the upper limit of the integration $L_\gamma^{\text{max}}=10^{52}~\text{ erg/s} $ in \citet{Ajello:2013lka} well exceeds the most luminous blazar observed, but it is consistent with the fact that the region of $L_{\gamma}>10^{50}~\text{ erg/s}$ contains no blazars and essentially no contribution to the $\gamma$-ray flux, as is shown in \figu{distro}.

These two populations of BL Lacs are characterized by different cosmological evolutions: the low-luminosity BL Lacs have a negative evolution, \ie, many of these objects are nearby, while high-luminosity BL Lacs have a positive evolution, \ie, are more abundant at high redshifts. Following the parametrization by Ajello et al. we find that the evolution changes from negative to positive when the luminosity is equal to $L_\gamma^{\text{evo}} \simeq 3.5 \times 10^{45}~\text{erg/s}$. A summary of these characteristics is reported in \Tab~\ref{tab:blazar}, together with the relative contributions from different blazar classes to the resolved and unresolved $\gamma$-ray flux.

In this work we focus on astrophysical neutrinos, whereas the distribution by Ajello et al. was originally created to study the $\gamma$-ray emission. Indeed using this distribution, it is possible to evaluate the $\gamma$-ray flux expected from BL Lacs and FSRQs. 
As stated before, a non-negligible fraction of this $\gamma$-ray flux can be produced by unresolved sources. Therefore it is quite natural to expect a similar behavior for neutrinos, and this aspect will be discussed in depth in the next sections. Moreover, if the contribution of unresolved sources to the neutrino flux were large enough, the lack of correlations between neutrinos and known $\gamma$-ray sources would become less problematic.

\begin{table*}[t]
\caption{Summary of some characteristics of BL Lacs and FSRQs: the evolution, the number of resolved and unresolved BL Lacs and FSRQs, and the resolved versus unresolved $\gamma$-ray flux contribution, listed separately for each population.}
\begin{center}
\begin{tabular}{llrrrr}
\hline
& \textbf{Evolution} &  \textbf{\# resolved sources} & \textbf{\# unresolved sources} & \textbf{Resolved flux} & \textbf{Unresolved flux} \\
\hline
\textbf{Low-luminosity BL Lacs} & Negative & 359 & 6070 & 64\% & 36\% \\
\textbf{High-luminosity BL Lacs} & Positive & 609 & 981 & 90\% & 10\% \\
\textbf{FSRQs} & Positive & 566 & 601 & 97\% & 3\% \\
\textbf{All blazars} & ---& 1534 & 7652 & 88\% & 12\% \\
\hline
\end{tabular}
\end{center}
\label{tab:blazar}
\end{table*}%

\subsection{The expected flux of astrophysical neutrinos}
\label{sec:neutrinoflux}

The expected flux of astrophysical neutrinos produced by blazars can be determined using the neutrino flux from a single source (identified by its luminosity and its redshift) and the distribution of blazars in the universe. 

The neutrino flux at Earth $\Phi_s$ produced by a single source with $\gamma$-ray luminosity $L_\gamma$ at redshift $z$ is given by the following expression:

\begin{equation}
\frac{d\Phi_s}{dE}(\ell,z,E,\xi(\ell))= \frac{1}{4 \pi D_c(z)^2}\left[ \frac{1}{E}\frac{dL_\nu}{dE}(\ell,E(1+z))\right] \times \xi(\ell),
\label{eq:nufluxsingle}
\end{equation}
where $\xi(\ell)$ is the baryonic loading of the specific source, the neutrino luminosity spectra $E dL_\nu/dE$ are represented in \figu{neutrinos_sequence} for a baryonic loading  $\xi=1$ and $D_c(z)$ is the comoving distance, defined in Appendix \ref{sec:ajello}. The relation between the total neutrino luminosity of the blazar and its $\gamma$-ray luminosity is given by \equ{neutrino_luminosity}. 

The cumulative flux of neutrinos at Earth is given by convolving the single-source neutrino flux with the distribution of BL Lacs and FSRQs over $L_\gamma$ and redshift $z$, as follows:

 \small
 \begin{equation}
 \begin{split}
\frac{ d\Phi_{\text{tot}}}{dE} (E,\xi(\ell))=&\left[ \int_{z_1}^{z_2} \int_{\ell_1}^{\ell_2} \frac{dN}{dz d\ell} \times \frac{d\Phi_s}{dE}(\ell,z,E,\xi(\ell)) \ dz d\ell \right]_{\text{BL Lacs}}\\
 + &\left[ \int_{z_1}^{z_2} \int_{\ell_1}^{\ell_2} \frac{dN}{dz d\ell} \times \frac{d\Phi_s}{dE}(\ell,z,E,\xi(\ell)) \ dz d\ell \right]_{\text{FSRQs}},
 \end{split}
 \end{equation}
\normalsize
where the integration limits $\ell_1$, $\ell_2$, $z_1$ and $z_2$ are given in \append{ajello}. On the other hand, to calculate the contribution of resolved and unresolved blazars separately, the $\gamma$-ray luminosity must be integrated either in the range $[\ell_1,\ell_{\rm vis}]$ (unresolved) or $[\ell_{\rm vis},\ell_2]$ (resolved), where $\ell_{\rm vis}$ is the average luminosity threshold discussed in section \ref{sec:sourcedist}, corresponding to a flux of $\phi_\gamma^{\rm vis}=4\times 10^{-12}~{\rm erg~cm^{-2}~s^{-1}}$. Therefore $\ell_{\rm vis}$ is a function of the redshift, namely $\ell_{\rm vis}(z)=4\pi \times \phi_\gamma^{\rm vis} \times D_c(z)^2 (1+z)^2$. 

It is important to remark that in the previous procedure it is implicitly contained the hypothesis that a source emits \textit{always} an average luminosity $L_\gamma$. 
  


\subsection{Neutrinos from the blazar TXS 0506+056}
\label{sec:txsintro}

In September 2017 a high-energy neutrino, event IceCube-170922A, was detected by IceCube \citep{IceCube:2018cha}. \textcolor{black}{Based only on its energy and direction, this event has a probability of 50\% to be an astrophysical neutrino rather than have atmospheric origin. On the other hand, after the neutrino detection} different $\gamma$-ray telescopes have detected a flaring state of the source TXS 0506+056 \citep{IceCube:2018dnn}, which lies in a direction consistent with that of the IceCube event. \textcolor{black}{This correlation in space and time increased the probability of this event being of astrophysical origin, excluding the atmospheric hypothesis at the $\sim 3.5\sigma$ level. }

\textcolor{black}{The aforementioned astrophysical source is considered to be a BL Lac, located} at redshift $z_{\scalebox{.8}{$\scriptstyle \text{TXS}$}}=0.3365 \pm 0.0010$, with a luminosity of $L_\gamma \simeq 1.3 \times 10^{47}~\text{erg/s}$ during the weeks before and after the neutrino event \citep{IceCube:2018dnn}. The luminosity averaged over all historical data is $L_\gamma \simeq 2.8 \times 10^{46}~\text{erg/s}$~\citep{IceCube:2018dnn}, as marked in the right panel of \figu{distro}. From here on we will use this time-averaged luminosity to study the characteristics of TXS 0506+056 within the blazar population. 

This discovery represents a breakthrough in the field of multi-messenger astronomy, since it is the first evidence of correlation between high-energy neutrinos and $\gamma$-rays.
Several theoretical papers have been written after this discovery attempting to explain the neutrino emission with blazar radiation models \citep{Ahnen:2018mvi,Gao:2018mnu,Keivani:2018rnh,Padovani:2018acg,Cerruti:2018tmc,Liao:2018pta}. It is therefore important to analyze the role of this source in the context of our general description, which focuses on the entire blazar population.
Note that, looking back at historical data, IceCube has also found an excess from the same direction between September 2014 and March 2015, with respect to atmospheric backgrounds. This constitutes $3.5 \sigma$ evidence of neutrino emission from the direction of TXS 0506+056, independent and prior to the 2017 flaring episode \citep{IceCube:2018cha}. On the other hand, the FSRQ PKS 0502+049 displayed flaring activity around the time of the neutrino flare observed by IceCube between 2014--2015 flare, and its position is compatible with the angular resolution of the events~\citep{Padovani:2018acg, He:2018snd}, which makes this source another interesting test case for our model. In \Secs~\ref{sec:txs} and~\ref{sec:pks} we discuss the implications of the current work to both these sources.

\begin{figure}[t]
\centering
\includegraphics[width=0.48\textwidth,angle=0]{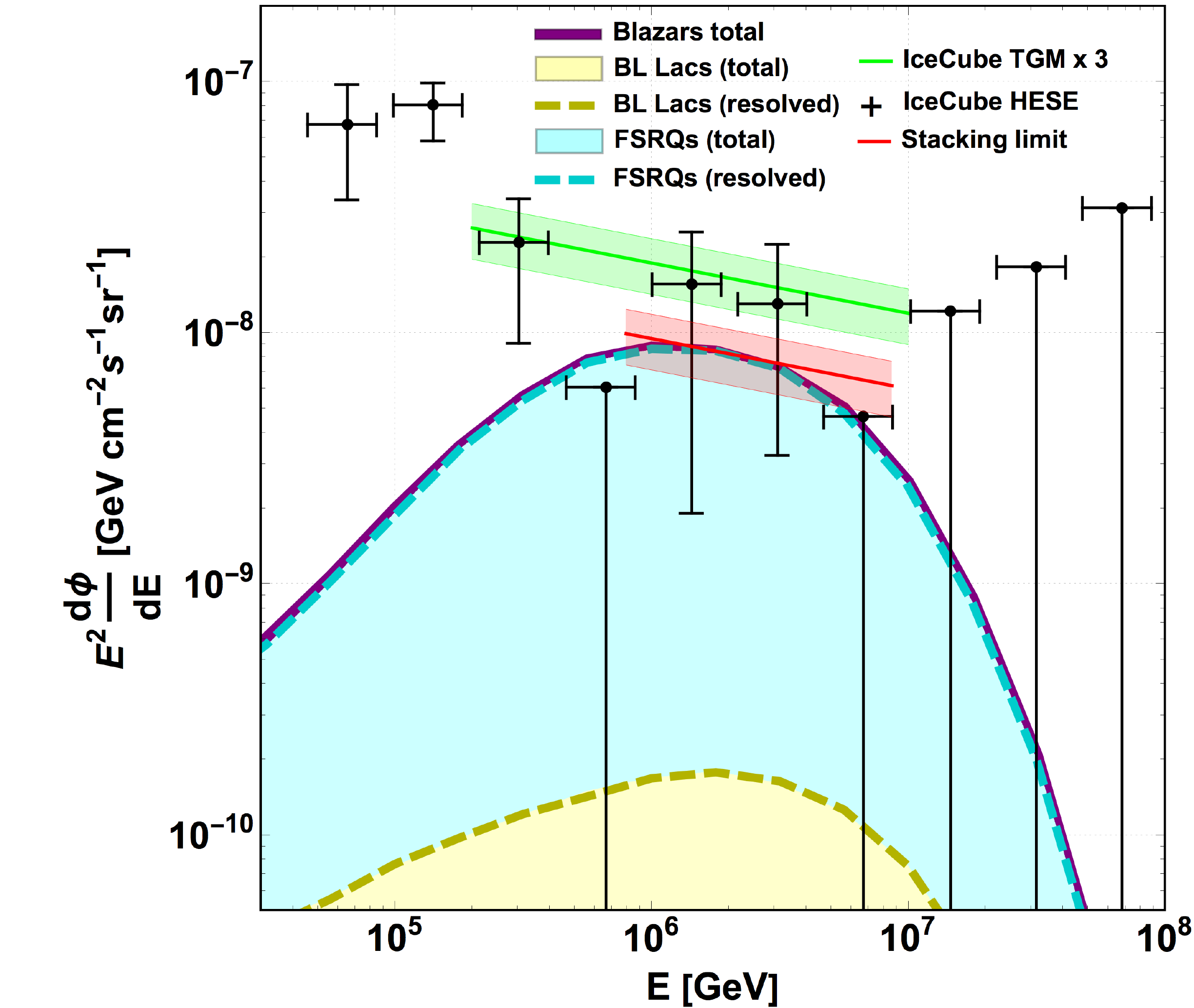}
\caption{Scenario 1: constant baryonic loading for all sources. The black points represent the IceCube HESE (all flavor, \citet{Aartsen:2015knd}), the green line represents the flux of throughgoing muons (x3, \citet{icemuon}) and the red line represents the current IceCube blazar stacking limit~\citep{stackbl}.
The shaded yellow and cyan regions denote the contribution of BL Lacs and FSRQs, respectively, to the total flux of neutrinos, whereas the dotted yellow and cyan curves denote the contribution from resolved sources only. The purple solid curve represents the total neutrino flux expected from blazars. In this case most of  the flux is powered by resolved sources, particularly FSRQs.}
\label{fig:scenario1}
\end{figure}

\begin{figure*}[t]
\includegraphics[width=0.31\textwidth]{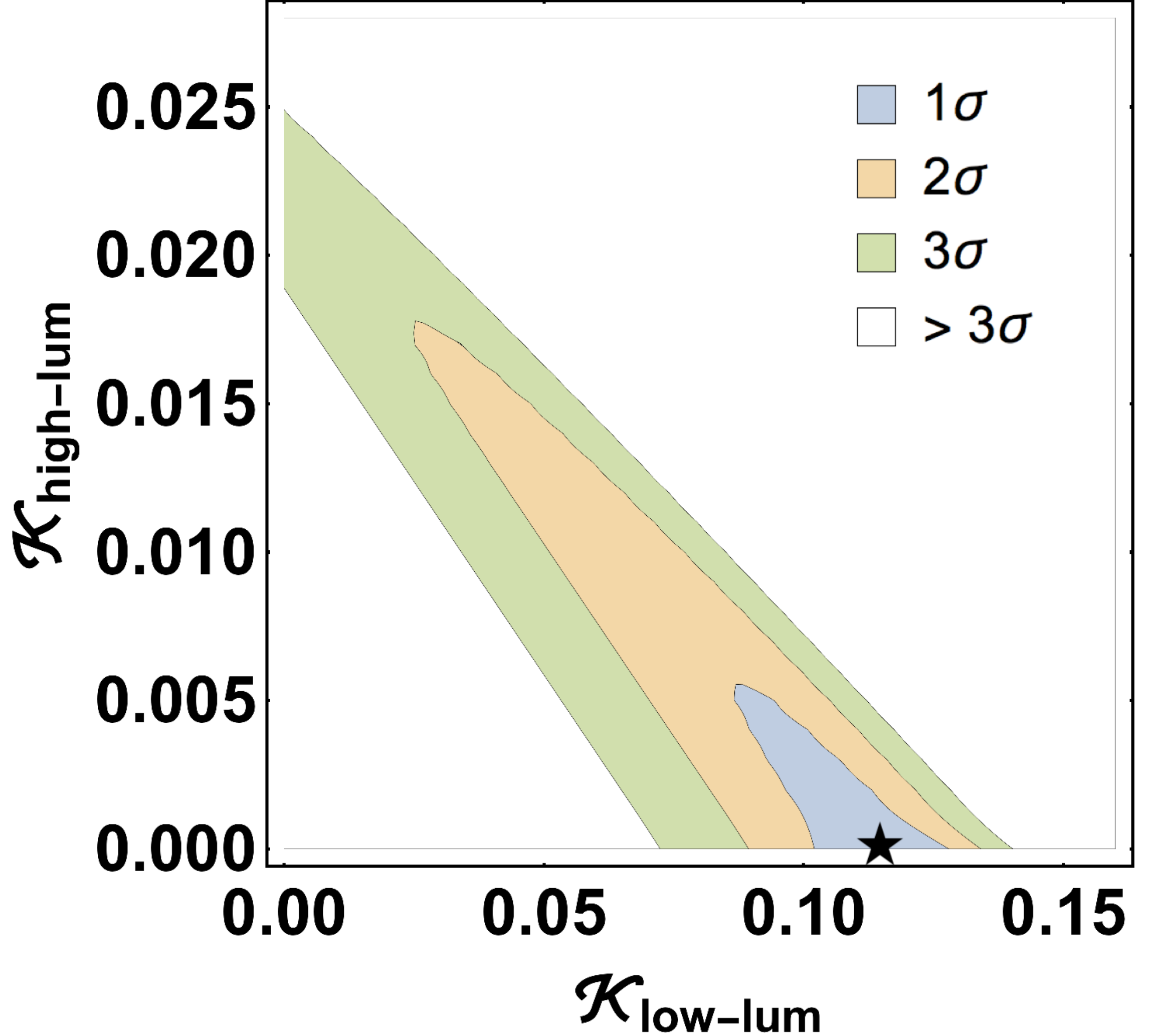}
\includegraphics[width=0.32\textwidth]{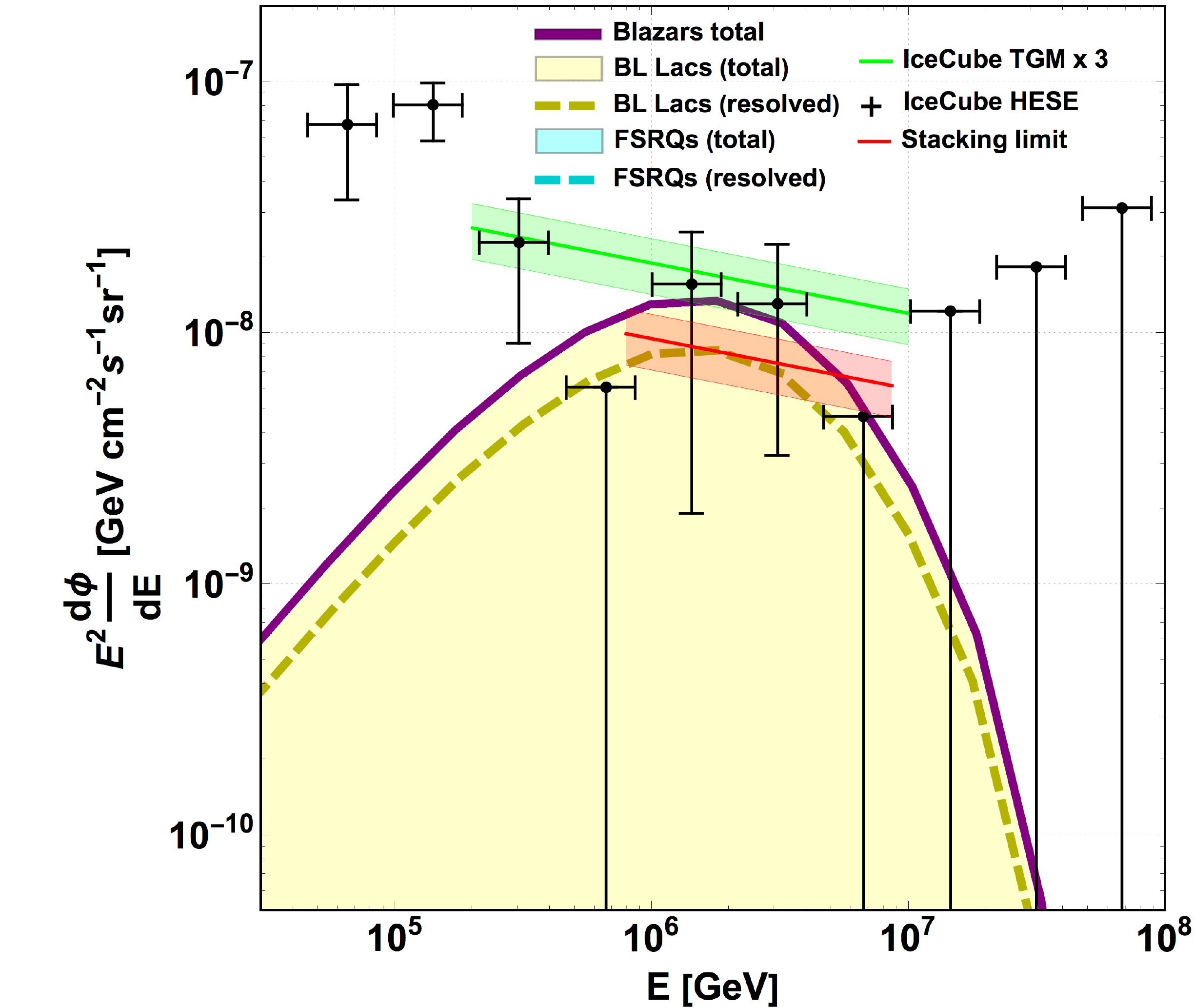}
\includegraphics[width=0.32\textwidth]{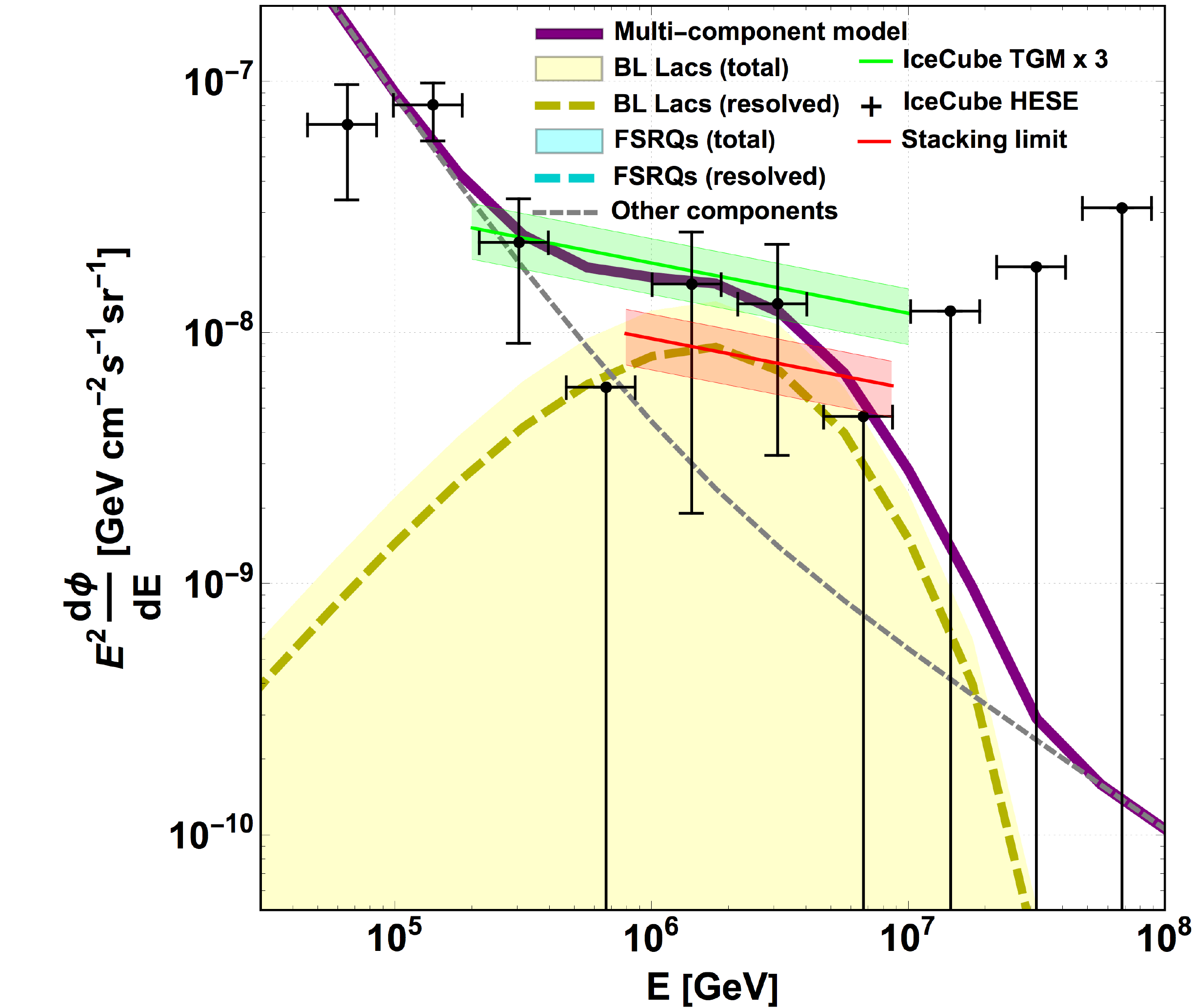}
\caption{Scenario 2: constant ratio $\mathcal{K}=L_\nu/L_\gamma=\epsilon_\nu \times \xi$. 
Left: scan of the values of $\mathcal{K}=\epsilon_\nu \times \xi$ for low- and high-luminosity sources, taking into account both IceCube observations and the stacking limit. The number of standard deviations $\sigma$ is calculated for 2 degrees of freedom. The best fit is marked with a black star.
Middle: theoretical expectation for the neutrino flux, as described in \figu{scenario1}, compared to the IceCube results. The flux of neutrinos is obtained considering the best fit values of $\epsilon_\nu \times \xi$, \ie, $\epsilon_\nu \times \xi=10.5 \%$ for low-luminosity sources and $\epsilon_\nu \times \xi=0 \%$ for high-luminosity sources. 
In this scenario, the neutrino flux is powered by both unresolved and resolved sources. Right: the same as the middle panel, including the effect of other possible contributions (multi-component model), such as the residual atmospheric background and Galactic neutrinos, as discussed in \citet{pal3}.} 
\label{fig:scenario2}
\end{figure*}

\section{Results}
\label{sec:res}
\textcolor{black}{In this section we analyze the consequences of the hypothesis that the most energetic neutrinos observed by IceCube are produced by blazars. This means that the neutrino flux at the highest energies is saturated by blazars, while at lower energies additional contributions may be present due to the expected peaked nature of the blazar neutrino flux.
Although other studies \citep{stackbl,Murase:2018iyl,Hooper:2018wyk} disfavor blazars as the dominant source of the astrophysical neutrinos  based on different arguments, it is still premature to rule out this source class in light of the potential contribution from unresolved blazars and the fact that they may dominate only at the highest energies. One motivation for such an investigation is that blazars dominate the gamma-ray sky above 100 GeV; if these gamma rays are of hadronic origin, the same sources ought to efficiently produce neutrinos.}

We test three different hypotheses for the baryonic loading and therefore for the relation between the the $\gamma$-ray luminosity and the neutrino luminosity defined in \equ{neutrino_luminosity}:
\begin{description}
 \item[Scenario 1: Constant baryonic loading] We assume a constant baryonic loading $\xi(\ell) = \tilde{\xi}$ for all sources, as in \citet{Zhang:2016vbb};
 \item[Scenario 2: Constant ratio $\boldsymbol{L_\nu/L_\gamma}$] We assume that the ratio between neutrino luminosity and $\gamma$-ray luminosity is constant, \ie, $L_\nu/L_\gamma \equiv \mathcal{K} = const$. This assumption is model-independent and has been used in previous works, such as~\citet{Kadler:2016ygj,righi,Halzen:2016uaj,Wang:2015woa} to evaluate the flux of neutrinos from BL Lacs. In the context of our model, it implies that the product $\xi \times \epsilon_\nu=\mathcal{K}$ (\cf, \equ{neutrino_luminosity}) is constant, which means that $\xi \propto (\epsilon_\nu)^{-1}$. We allow for  different values for low-luminosity BL Lacs and high-luminosity BL Lacs/FSRQs (for the purposes of this analysis we consider high-luminosity sources as one group), reflecting the potentially different baryonic loadings. 
 \item[Scenario 3: Baryonic loading evolving with $\boldsymbol{L_\gamma}$] We assume that the baryonic loading changes continuously as a function of $\ell$; this function is assumed to be universal for BL Lacs and FSRQs.
This is a generalization of the second scenario. 
\end{description}
 
The theoretical predictions will be compared with the through-going muon flux~\citep{icemuon}, with the high-energy starting events~\citep{Aartsen:2015knd} above 100 TeV and with the blazar stacking limit~\citep{stackbl}. As discussed in depth in~\citet{pal1,pal2,pal3}, the IceCube data below 100 TeV can be affected by the presence of Galactic neutrinos and residual atmospheric background (both conventional and prompt neutrinos). For this reason, we choose the throughgoing muons as reference for the extragalactic neutrino flux.

We now discuss the results obtained in the three scenarios described above.

\subsection{Scenario 1: Constant baryonic loading}

The diffuse flux obtained choosing a constant baryonic loading is represented in \figu{scenario1}, where a baryonic loading $\tilde{\xi}=150$ has been chosen in order to match and not overshoot the present IceCube stacking limit for blazars~\citep{stackbl}. Let us remark that the stacking limit has been obtained by IceCube assuming a power law spectrum, although we expect a different spectrum for blazars. A new preliminary analysis has been presented in the conference proceedings~\citet{Aartsen:2017mau} based on different $p\gamma$ models. However, these models are not consistent with each other and yield very different predictions. We therefore consider in this work the only published stacking limit, choosing that obtained for a spectral index of 2.2, which is roughly that of the IceCube throughgoing muons.
In \figu{scenario1}, the cumulative flux is represented using a purple solid curve, whereas the contributions from BL Lacs and FSRQs are shown using the shaded yellow and cyan regions, respectively. From the same plot we can also see the flux produced by resolved sources (dashed curves), which in this case is not distinguishable from the total flux produced by BL Lacs and FSRQs. This implies that:
\begin{quote} 
\textit{Assuming a constant baryonic loading, the flux of neutrinos is fully powered by resolved sources. Therefore, it is not possible to simultaneously interpret the IceCube observations and obey the stacking limit.}
\end{quote}
A higher baryonic loading would produce a higher neutrino flux, improving the agreement with the observations but violating the stacking limit. Therefore, assuming that astrophysical neutrinos are produced by blazars, a constant baryonic loading is not a viable assumption to interpret the astrophysical neutrinos. Even if one allowed for two different constant values  for BL Lacs and FSRQs, the result would not change because the neutrino flux is  determined by the resolved sources; therefore, it is not possible to saturate the IceCube measurement obeying the stacking limit at the same time. 

\subsection{Scenario 2: Constant ratio $L_\nu/L_\gamma$}
\label{sec:const}

As a second scenario, we consider a constant  ratio $\mathcal{K} \equiv L_\nu/L_\gamma$.  We assume two different values for low-luminosity BL Lacs ($\mathcal{K}_{\text{low-lum}}$) and high-luminosity BL Lacs/FSRQs ($\mathcal{K}_{\text{high-lum}}$).  This assumption comes from the following consideration: in order to power the neutrino flux detected by IceCube without violating the stacking limit, the contribution of unresolved sources (mainly low-luminosity sources) has to be enhanced and, at the same time, the contribution of resolved sources (mainly high-luminosity sources) has to be suppressed. As a splitting point, we choose $L_{\gamma}^{\text{evo}}$, which possibly separates (for BL Lacs) two different populations; see right panel of \figu{distro} and \Sec~\ref{sec:sourcedist}. Therefore we have:
$$
L_\nu=
\left\{
\begin{array}{cc}
\mathcal{K}_{\text{low-lum}} \, L_\gamma & \mbox{for} \ \ \ L_\gamma \leq L_{\gamma}^{\text{evo}} \\
\mathcal{K}_{\text{high-lum}} \, L_\gamma & \mbox{for} \ \ \ L_\gamma > L_{\gamma}^{\text{evo}}
\end{array}
\right.\
$$
where $L_{\gamma}^{\text{evo}}=3.2\times 10^{45}~\text{erg/s}$.

Since in scenario 2 the value of $\mathcal{K}=\epsilon_\nu \times \xi$ is constant for all sources, for our source model this implies that the baryonic loading scales as $\xi(\ell) \propto (\epsilon_\nu(\ell))^{-1}$,
\ie, the baryonic loading $\xi(\ell)$ decreases with luminosity in a way that it exactly compensates for the increasing  neutrino production efficiency $\epsilon_\nu(\ell)$. Note that the behavior of the baryonic loading depends on the particular model of neutrino production in the source, while the assumption $L_\nu \propto L_\gamma$ is not model dependent.

We compute the total and resolved fluxes from BL Lacs and FSRQs and we compare them with the measured through-going muon flux~\citep{icemuon} and with the present stacking limit~\citep{stackbl}, for each value of $\mathcal{K_{\text{low-lum}}}$ and $\mathcal{K_{\text{high-lum}}}$. The scan of these two parameters is represented in the left panel of \figu{scenario2}, where we show the $1\sigma$, $2\sigma$, $3\sigma$, and $\geq 3\sigma$ regions. The best fit is given by the following set of parameters:
\begin{equation}
\mathcal{K}_{\text{low-lum}} = 10.5 \% ;  \ \ \ \mathcal{K}_{\text{high-lum}} = 0 \%.
\end{equation}
The interpretation of that result is that the neutrino flux must be powered by low-luminosity sources in order to avoid overshooting the stacking limit. Within 1$\sigma$, $\mathcal{K}_{\text{high-lum}} \leq 0.5 \%$ is allowed, providing an upper limit to the contribution of high-luminosity sources.

Within scenario~2, we can interpret the IceCube data at the highest energies without violating the stacking limit, as illustrated in the middle panel of \figu{scenario2}. From this plot we notice that the contribution of FSRQs is absent while BL Lacs (particularly those with low luminosity) provide the dominant contribution to the diffuse neutrino flux (yellow shaded region). Combining this result with the idea of the multiple components  presented in \citet{pal3}, which may power the neutrino flux at lower energies, we obtain the total neutrino flux (shown in the right panel of \Fig~\ref{fig:scenario2}), as combination of \textit{i)} the neutrino flux from blazars plus \textit{ii)} the neutrino flux coming from residual background (atmospheric neutrinos, $\phi(E) \sim E^{-3.7}$ below 100 TeV) and Galactic neutrinos ($\phi(E) \sim E^{-2.6}$ with energy cutoff at 150 TeV); we denote these additional neutrinos as \lq\lq other components\rq\rq. The 
additional component has not been re-fitted in this paper, but it is exactly the same reported in \citet{pal3}. 
It is important to remark that in principle the atmospheric background should have been already subtracted in the IceCube data points. On the other hand, in \citet{pal3} we discuss the possibility that a certain residual background can still affect the IceCube measurement, even after the subtraction of the background (see particularly Fig.~2 of \citet{pal3}.). The residual atmospheric background may become relevant especially below 100 TeV in the high-energy starting events analysis. We notice from the right panel of \figu{scenario2} that the combination of scenario 2 and these additional components allows for an interpretation of the shape of the spectrum measured by IceCube.

\textcolor{black}{The result of scenario 2 is much more optimistic than that of \citet{Hooper:2018wyk}. The main reason is that \citet{Hooper:2018wyk} only consider the contribution of detected sources, presenting an updated version of the analysis contained in \citet{stackbl}. On the contrary, we have shown that under the hypotheses of scenario 2, the contribution of unresolved sources can be relevant.}


\subsection{Scenario 3: Baryonic loading evolving with luminosity as a power law}
\label{sec:barload}

Here we generalize the previous scenario and assume that the baryonic loading scales with the luminosity $\ell$ as a continuous function, defined as follows: 
\begin{itemize}
\item for low-luminosity sources, we roughly replicate $\mathcal{K}_{\text {low-lum}} \equiv L_\nu/L_\gamma = 10.5 \%$ suggested by the best fit result of scenario~2. Therefore the baryonic loading is given by $\xi(\ell) \simeq \mathcal{K}_{\text{low-lum}}/\epsilon_\nu(\ell)$ for our production model;
\item for high luminosity sources we use the information $\mathcal{K}_{\text{high-lum}} \equiv L_\nu/L_\gamma < 0.5 \% $ from scenario~2, and we derive an upper limit for the baryonic loading, indicated above;
\item for the region in between, we use the information from TXS 0506+056 in the model presented in \citet{Gao:2018mnu}, where the baryonic loading of the source TXS 0506+056 is calculated to be $\xi \simeq 3\times 10^4$ during the flare. 
\end{itemize}
The information provided above can be checked looking at the middle panel of \figu{eddington}, where the function $\mathcal{K}$ is represented as a function of luminosity. Let us note that at low luminosity the value of $\mathcal{K}$ for FSRQs is not equal to $\simeq 10\%$, as stated before. On the other hand we have to take into account that at low luminosity there are no FSRQs at all (see right panel of \figu{distro}); therefore this fact cannot produce any effect on the calculation.

The inclusion of the information from TXS 0506+056 is relevant for our purpose, since TXS 0506+056 has been identified as the first possible source of an IceCube neutrino event \citep{IceCube:2018cha,IceCube:2018dnn}. Although it may not be representative for the whole population, it is so far the only direct evidence of a correlation between neutrinos and $\gamma$-rays, and therefore it is the best available piece of information at this point. As a consequence, our model is consistent with TXS 0506+056 by construction.

\begin{figure}[t]
\includegraphics[width=0.49\textwidth,angle=0]{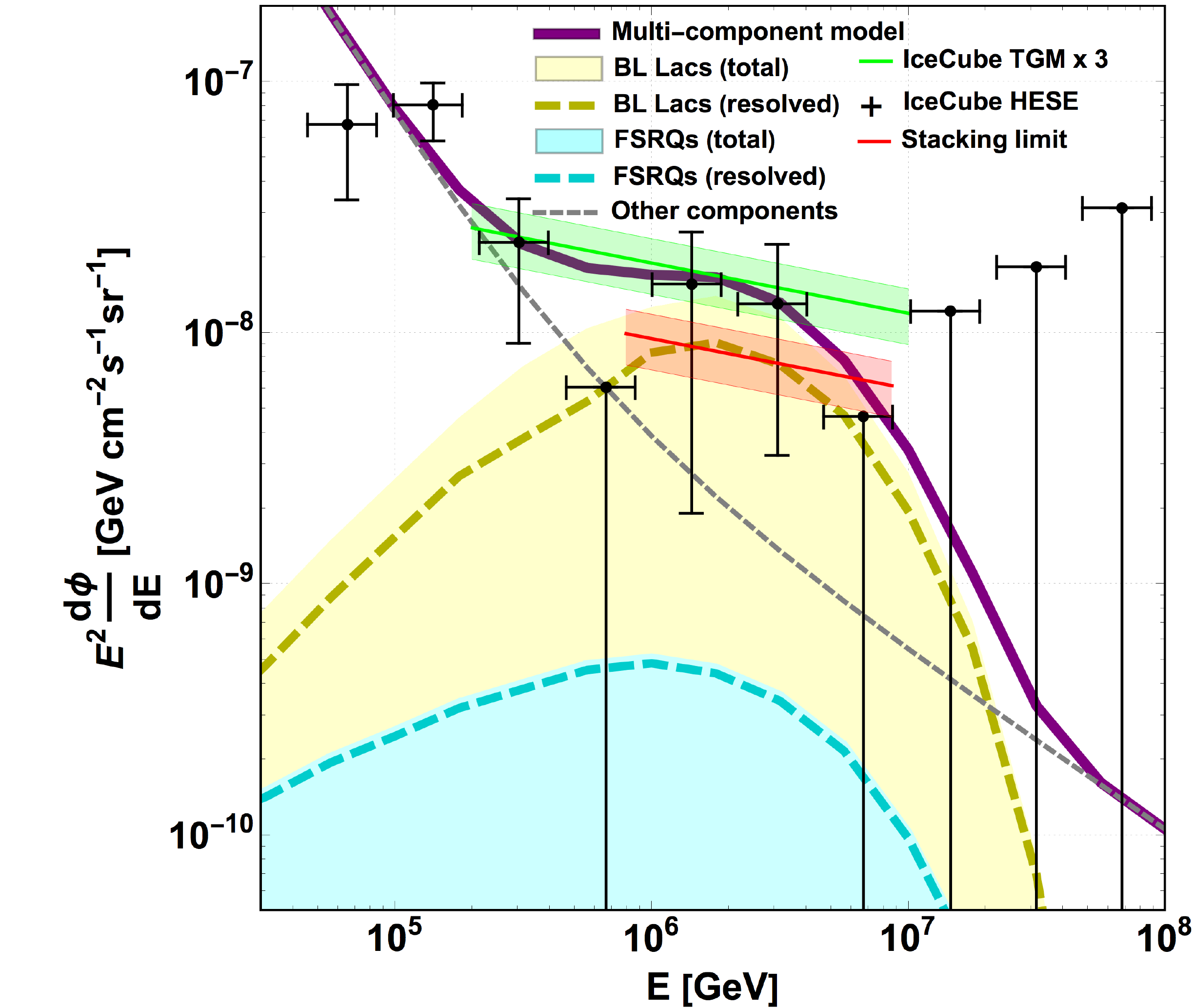}
\caption{Scenario 3: baryonic loading changing continuously with luminosity, evolving from a baryonic loading larger than $10^5$ for low-luminosity sources to a baryonic loading smaller than 10 for high-luminosity sources, see main text. As in scenario 2, the neutrino flux is powered by both unresolved and resolved sources, with a dominant contribution from low-luminosity BL Lacs.}
\label{fig:scenario3}
\end{figure}

As in scenario 2,  we can describe the through-going muon flux without violating the stacking bound. The results and the main conclusion are similar to the ones obtained in the second scenario, as it can be observed in \Fig~\ref{fig:scenario3}. However, this scenario allows for a small contribution by FSRQs, which was not present in the last scenario.
The actual function obtained for the baryonic loading is represented by the blue curve in the left panel of \Fig~\ref{fig:eddington}.  On the other hand, in this case we have chosen the upper limit (within $1\sigma$) of the baryonic loading for high-luminosity sources; therefore the contribution of FSRQs can also be lower or even negligible, as can be seen by the blue shaded region in the left panel of \Fig~\ref{fig:eddington}, representing the uncertainty in the baryonic loading. Note that the TXS 0506+056 lies on the verge between low-luminosity sources, which dominate the diffuse neutrino flux, and the flux-limited high-luminosity sources; it is therefore a rather special case in the context of our model.

In conclusion, the result from the second and third scenarios is comparable: the contribution of high-luminosity blazars must be suppressed in order to not violate the stacking limit, whereas the contribution of low-luminosity blazars has to be dominant. This conclusively means that:
\begin{quote}
\textit{Low-luminosity BL Lacs can be bright neutrino sources and they should have a baryonic loading higher than $10^5$ in order to power the neutrino events in IceCube.} On the contrary, FSRQs must have a much lower baryonic loading (about 10) and are therefore dim in the neutrino channel.
\end{quote}
From here on this scenario will represent our baseline model and its implications will be discussed in the next section.

\begin{figure*}[tbp!]
\includegraphics[width=0.32\textwidth]{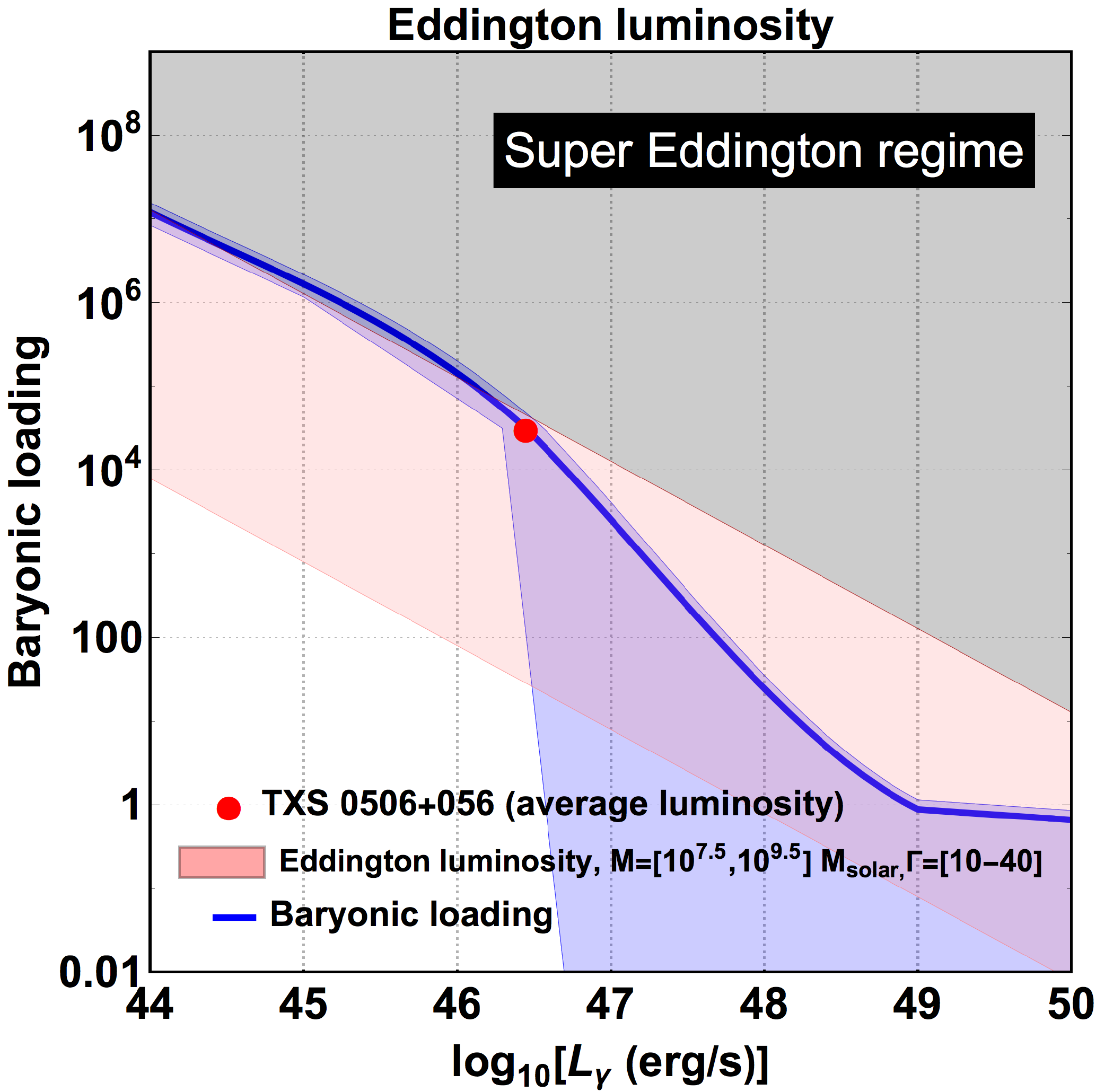}
\includegraphics[width=0.32\textwidth]{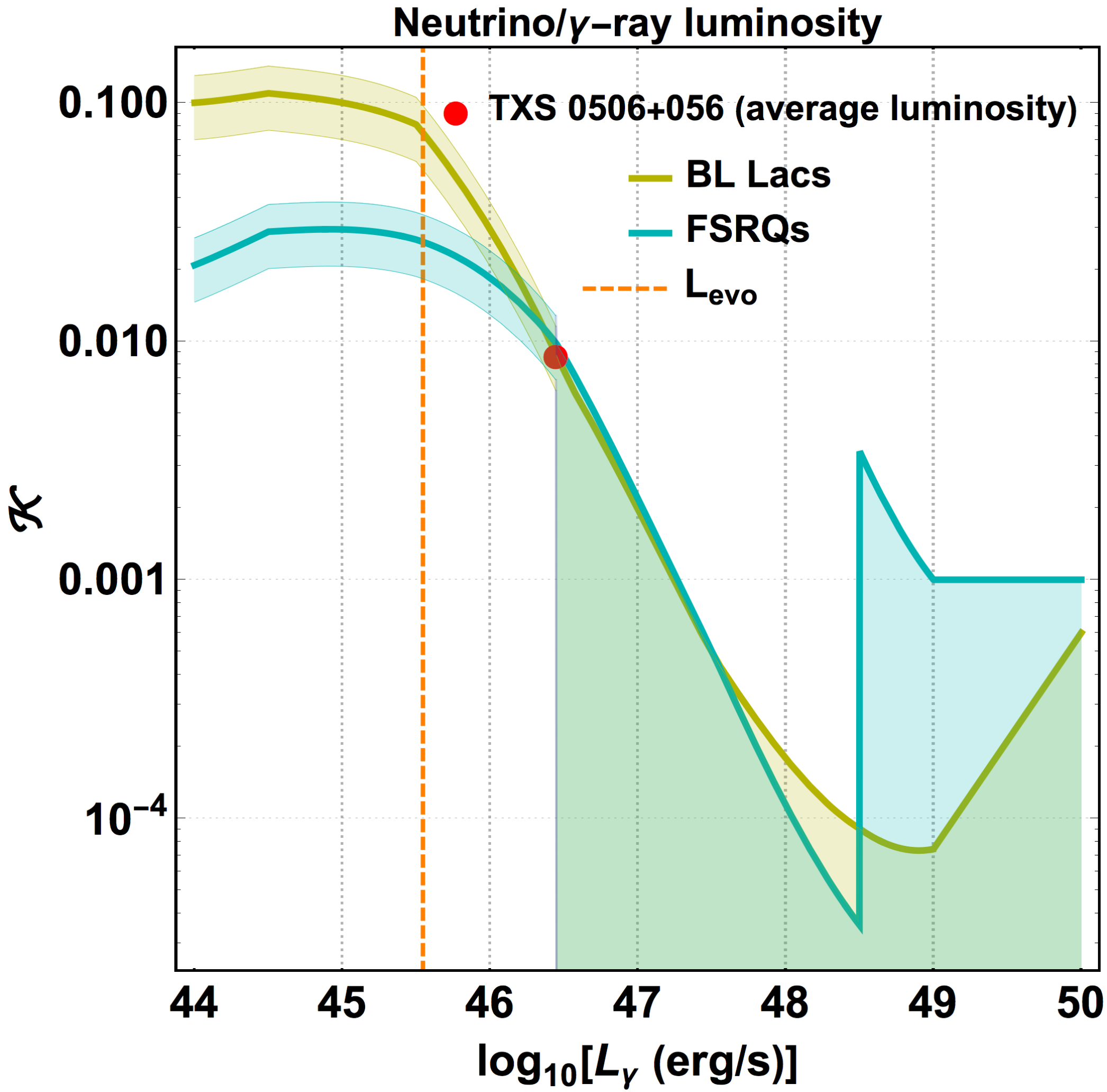}
\includegraphics[width=0.32\textwidth]{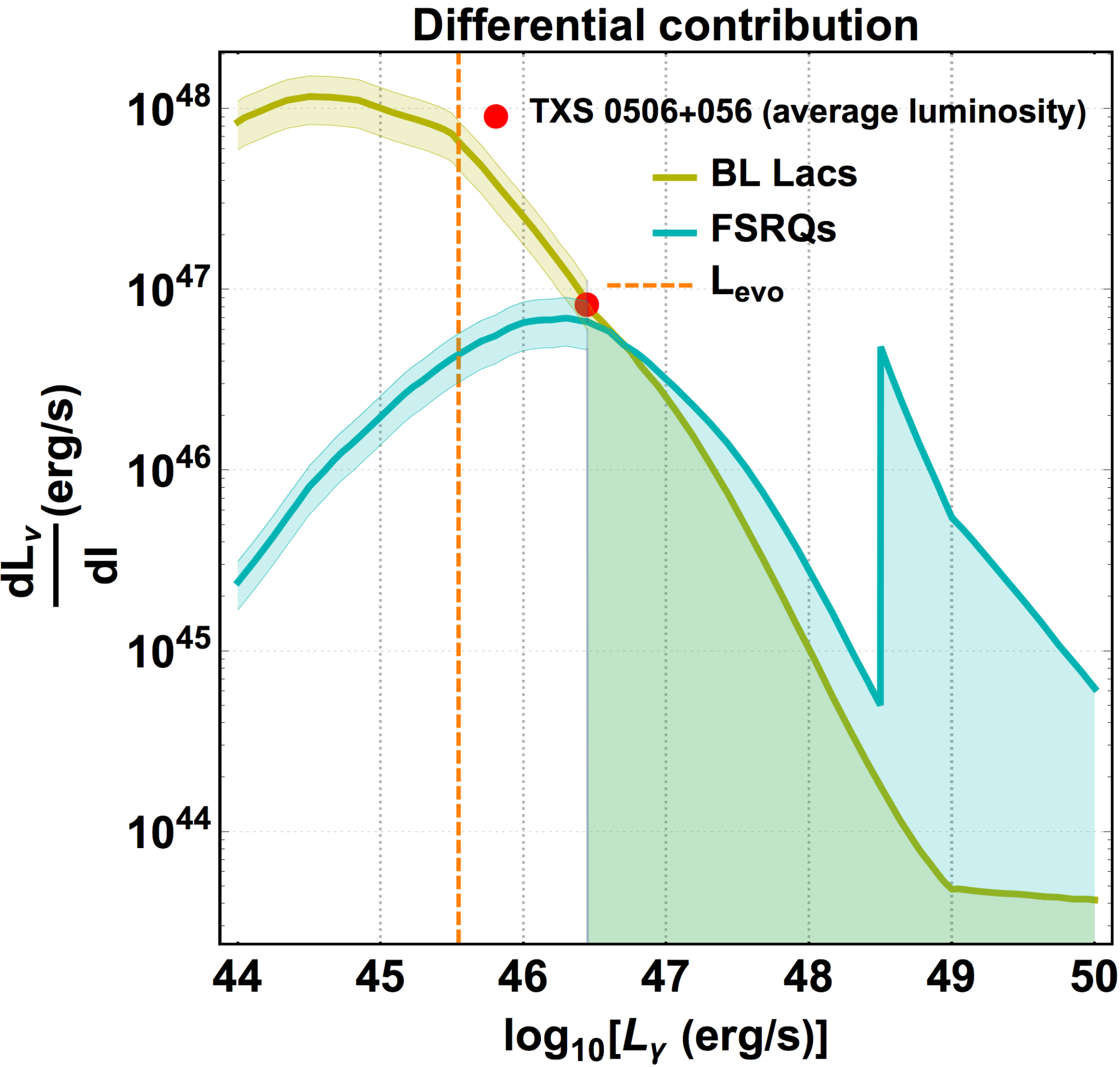}
\caption{Left: baryonic loading as a function of $L_\gamma$ (best-fit as blue curve, uncertainties represented by blue band). The region disfavored by the Eddington luminosity is represented in gray. The red region represents the Eddington luminosity according to a black hole mass range $10^{7.5}$-$10^{9.5}M_\odot$ with a Lorenz factor range of $\Gamma=10-40$. Middle: neutrino to gamma-ray luminosity as a function of $L_\gamma$. Right: same quantity as in the middle panel multiplied by $dN/d\ell$, the differential number of sources of luminosity $\ell$. All panels given for scenario~3. 
}
\label{fig:eddington}
\end{figure*}

\section{Discussion}
\label{sec:discussion}

In this section we discuss some aspects to test the plausibility of our baseline model (scenario~3), namely: \textit{i)} the evolution with $\gamma$-ray luminosity of the baryonic loading and neutrino luminosity of blazars, including a comparison of the baryonic loading with the Eddington luminosity; \textit{ii)} the discussion concerning the multiplet constraint, \ie, the absence of two neutrinos coming from the same source; \textit{iii)} the comparison between our general picture and the expectations from the blazar TXS 0506+056, which rises at present great interest due to the recent evidence for a correlation between one of its flares and one IceCube neutrino detected in September 2017; \textit{iv)} a discussion concerning the object PKS 0502+049 as neutrino emitter.

\subsection{Source parameters with $\gamma$-ray luminosity}

In order to test the plausibility of our result, we check if the baryonic loading that we obtain is compatible with the Eddington luminosity. 
The Eddington luminosity is the maximum luminosity that a body (such as a star) can achieve when there is balance between the outward radiation force and the inward gravitational force. For accreting black holes, such as AGNs, of mass $M$, it is given by:
\begin{equation}
 L_{\text{edd}} = \frac{4\pi G m_p c}{\sigma_{\text T}} M \simeq 1.26 \times 10^{38} \left(\frac{M}{M_\odot} \right)~{\rm erg/s}.\label{eq:eddington}
\end{equation}
where $G$ is the gravitational constant, $m_p$ is the proton mass and $\sigma_{\text T}$ is the Thomson cross section. For the Lorentz factor $\Gamma$, a large range of values has been used in the literature of blazar modeling, e.g.\citep{Boettcher:2013wxa}. Here we have chosen a conservative range between 10 and 40. A range of supermassive black hole masses of $M \in [10^{7.5}, 10^{9.5}] M_\odot$ is assumed, which is typical for blazars~(\citet{tavecchioedd1,tavecchioedd2}, see also \Fig~7 of~\citet{eddingtoncinesi}).
Note that the Eddington luminosity is not a hard limit on the expected luminosity of blazars and may be exceeded, especially during flares~\citep{Sadowski:2015ena}.

For a jet dominated by baryons, we can use the Eddington luminosity to estimate the maximal plausible baryonic loading by comparing to the physical jet luminosity in $\gamma$-rays (corrected by the beaming factor) as
\begin{equation}
\xi(\ell) \lesssim \frac{L_{\text{edd}}}{L_\gamma/(2\Gamma^2)} \, .
\label{equ:ledd}
\end{equation}
The comparison between the baryonic loading $\xi(\ell)$ obtained in \sect{barload} and the Eddington luminosity limit is shown in the left panel of \figu{eddington}, where both the best-fit (blue curve) and the uncertainties (blue band) are represented. From this plot, we notice that the  baryonic loading obtained is compatible with the Eddington luminosity for the Lorentz factor and the black hole masses that we have discussed. Note that the quantity $L_{\text{edd}}$ may also vary with $L_\gamma$, due to the fact that the black hole mass and the ejected luminosity are slightly correlated, as found in~\citet{eddingtoncinesi}. On the other hand, this behavior is contained in the uncertainties that we have used, \ie, within the red region in the left panel of \figu{eddington}; therefore this comparison is sufficient for the purposes of this study. 
\textcolor{black}{It is however important to remark that Eq.~(\ref{equ:ledd}) is a necessary, rather than sufficient, condition. For example, in unified schemes of radio-loud AGNs \citep{Urry:1995mg}, in which BL Lacs are associated with FR I radio galaxies with radiatively inefficient accretion flows that do not produce broad emission lines, we expect highly sub-Eddington accretion powers and similarly sub-Eddington jet powers \citep{Sikora:2006xz}. While these studies do not seem to be compatible with our findings, one may speculate that the baryonic injection comes during flares only, when the Eddington luminosity may in fact be exceeded~\citep{Gao:2018mnu}.}

In the middle panel of \figu{eddington} we show the ratio $L_\nu/L_\gamma$ for our baseline scenario~3.
The ratio $L_\nu/L_\gamma$ evolves from  approximately 10\% to 0.01\% for BL Lacs (yellow curve). FSRQs exhibit a similar behavior up to $L_\gamma \simeq 3\times 10^{48}~{\rm erg/s}$, where $L_\nu/L_\gamma$ strongly increases due to the increasing neutrino production efficiency because of external radiation fields becoming relevant. It is noteworthy that  $L_\nu/L_\gamma$ strongly decreases in the range relevant for TXS 0506+056, compared to lower luminosities. This observation will be relevant later on in this analysis.

The right panel of \figu{eddington} shows the differential contribution  to the neutrino flux, where the  function $dL_\nu/d\ell$ is obtained as follows:
 \begin{equation}
\frac{dL_\nu}{d\ell} = \int_0^{z_{\text max}} dz \ 10^\ell \epsilon_\nu(\ell) \times \xi(\ell) \times \frac{dN}{d\ell dz} \frac{1}{(1+z)^2} \, .
\label{eq:xiepsdndl}
\end{equation}
Here $z_{\text max}=6$ for both BL Lacs and FSRQs. Let us recall from \equ{neutrino_luminosity} that $10^\ell \epsilon_\nu(\ell) \times \xi(\ell)=L_\nu(\ell)$, where $10^\ell=L_\gamma$. The factor $1/(1+z)^2$ refers to the fact that frequency and energy are redshifted. 
It is clear from this plot that the most important contribution to the high-energy neutrino flux comes from low-luminosity blazars, especially BL Lacs.

The results obtained in the second and third scenarios also have effects on the SED modeling, since they suggest that low-luminosity objects are rich in hadrons, which means the second hump of the SED may be produced by $\pi^0$ decays. For intermediate-luminosity objects the situation is unclear and both hadronic and hybrid (lepto-hadronic) models may be sufficient to explain the $\gamma$-ray and neutrino emission at the same time. High-luminosity objects (such as FSRQs) have a low baryonic loading and, in principle, they could be in agreement with hybrid or even purely leptonic models, since the contribution of these sources to the neutrino flux can be negligible, without affecting the quality of the result, in both scenarios 2 and 3.

\subsection{Multiplet constraint}


In this section we discuss the multiplet constraint, refering to the constraint from the non-detection of two or more events from the same source. Namely, we discuss whether the fact that IceCube has not yet observed any multiplet can test our model. This aspect has been emphasized in \citet{Murase:2016gly,Murase:2018iyl}, where it was remarked that the absence of neutrino multiplets (or point sources) constrains the contribution of BL Lacs at the level of 10\% at 100 TeV and less below 100 TeV. Therefore, in that work blazars are excluded to be the major contributors to the IceCube signal. Our prediction is compatible with the one of \citet{Murase:2018iyl} at 100 TeV, since we also find a contribution of $\sim 10\%$. On the other hand, we are not interested in explaining the low-energy events, since it is still unclear to which point the background affects events below 100 TeV. We focus instead on the throughgoing muon flux, therefore on neutrinos above 200 TeV. Under this assumption the multiplet constraint does not pose a problem to the model, as shown in the remainder of this section.


Let us consider the 9186 sources coming from the distribution by Ajello et al. Each source will contribute with a certain weight to the total neutrino flux expected from blazars and the weight $w_i$ depends 
on the luminosity $L_\nu$ and on the redshift $z$, as follows:
$$
w_i = \frac{L_\nu^i}{D_c(z^i)^2 (1+z^i)^2}, 
$$ 
where the relation between $L_\nu$ and $L_\gamma$ has been discussed in \Secs~\ref{sec:const} and~\ref{sec:barload} and $D_c(z)$ is defined in Appendix \ref{sec:ajello}. 
Let us remark that more than 90\% of the neutrino flux from blazars is powered by BL Lacs with $L_\gamma \leq 3.5 \times 10^{45} \mbox{erg/s}$ (about 6300 sources), following our baseline model (scenario 3).

\begin{figure}[t]
\includegraphics[scale=0.4]{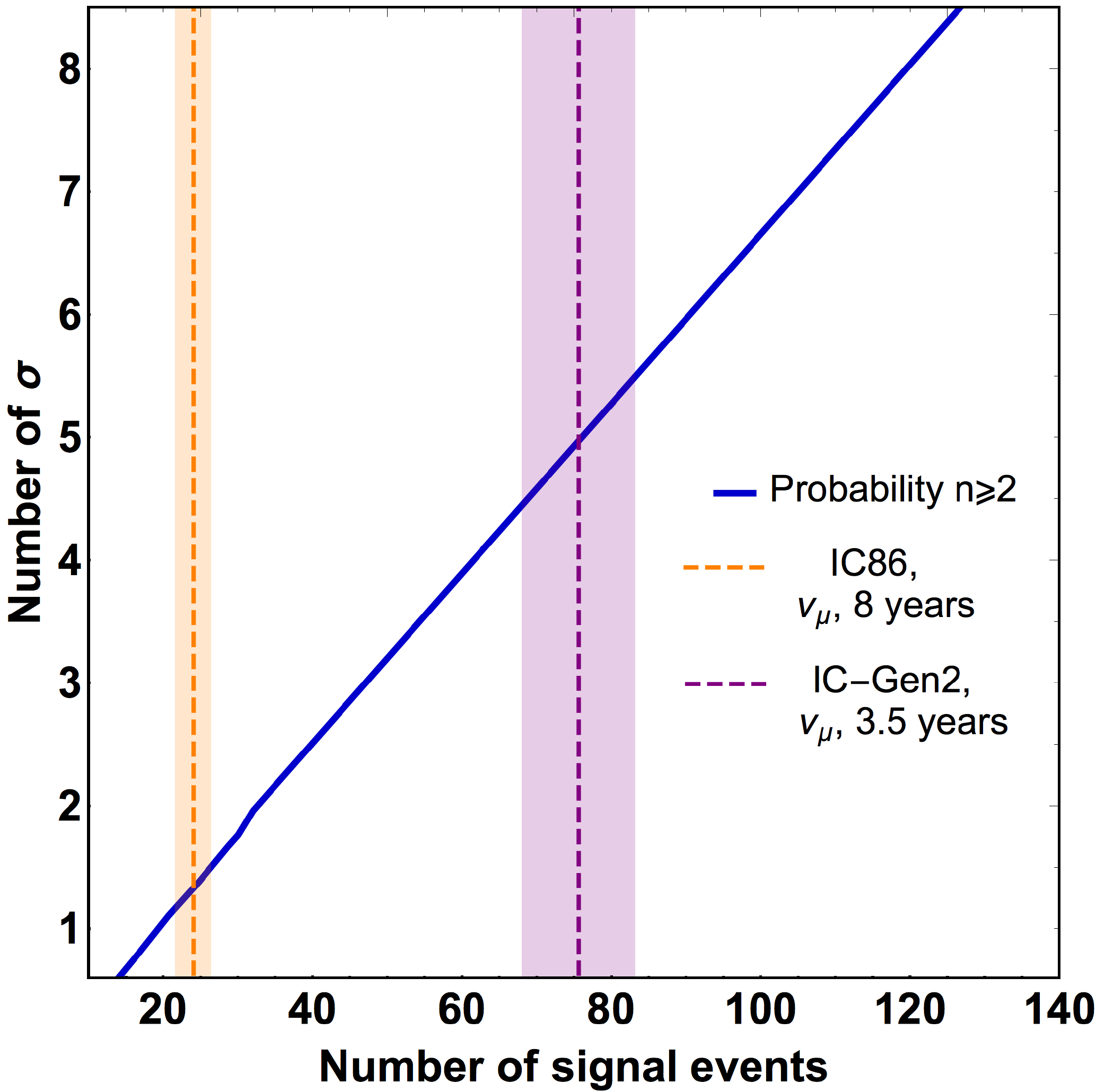}
\caption{Multiplet limit (number of standard deviations $\sigma$ of exclusion of the model due to the non-observation of multiplets) as a function of the observed number of signal events in the through-going muon dataset. In this analysis we use our baseline model (scenario 3) and we consider all 9186 blazars contained in the distribution by Ajello et al. 
We show as an orange band the present number of signal detected using throughgoing muons \citep{icemuon}. The purple band represents the number of years required for IceCube-Gen2 to reach the 5~$\sigma$ level \citep{Aartsen:2014njl}. The band is due to the uncertainty on the contribution of the atmospheric background (mainly prompt neutrinos) to the throughgoing muons, that is about 10\% (see Eq.12 of \citep{Palladino:2017aew}).} 
\label{plotmul}
\end{figure}

Using the previous information we can perform several simulations to evaluate the expected number of multiplets, given a certain number of observed signal events. Up to now, 36 throughgoing muons\footnote{In order to evaluate the expected number of multiplets we need events with good angular resolution, therefore we use the throughgoing muons dataset in this analysis since the average angular resolution is $\sim 1^\circ$.}  have been observed by IceCube  \citep{Aartsen:2017mau} and 2/3 of them are expected to be signal events, with an uncertainty of about 30\%. It translates in a tension between our model and the absence of multiplets of $1\sigma-1.7\sigma$, as it can be seen looking at the orange band of Fig.~\ref{plotmul}. Therefore, the absence of multiplets cannot be considered an issue for our model at this stage.


On the other hand, the next IceCube generation \citep{Aartsen:2014njl}, in which the exposure is expected to be 6-7 times larger than the present one, can test and eventually rule out our model in about 3.5 years (purple band).

\begin{figure*}[t]
\includegraphics[width=0.485\textwidth,angle=0]{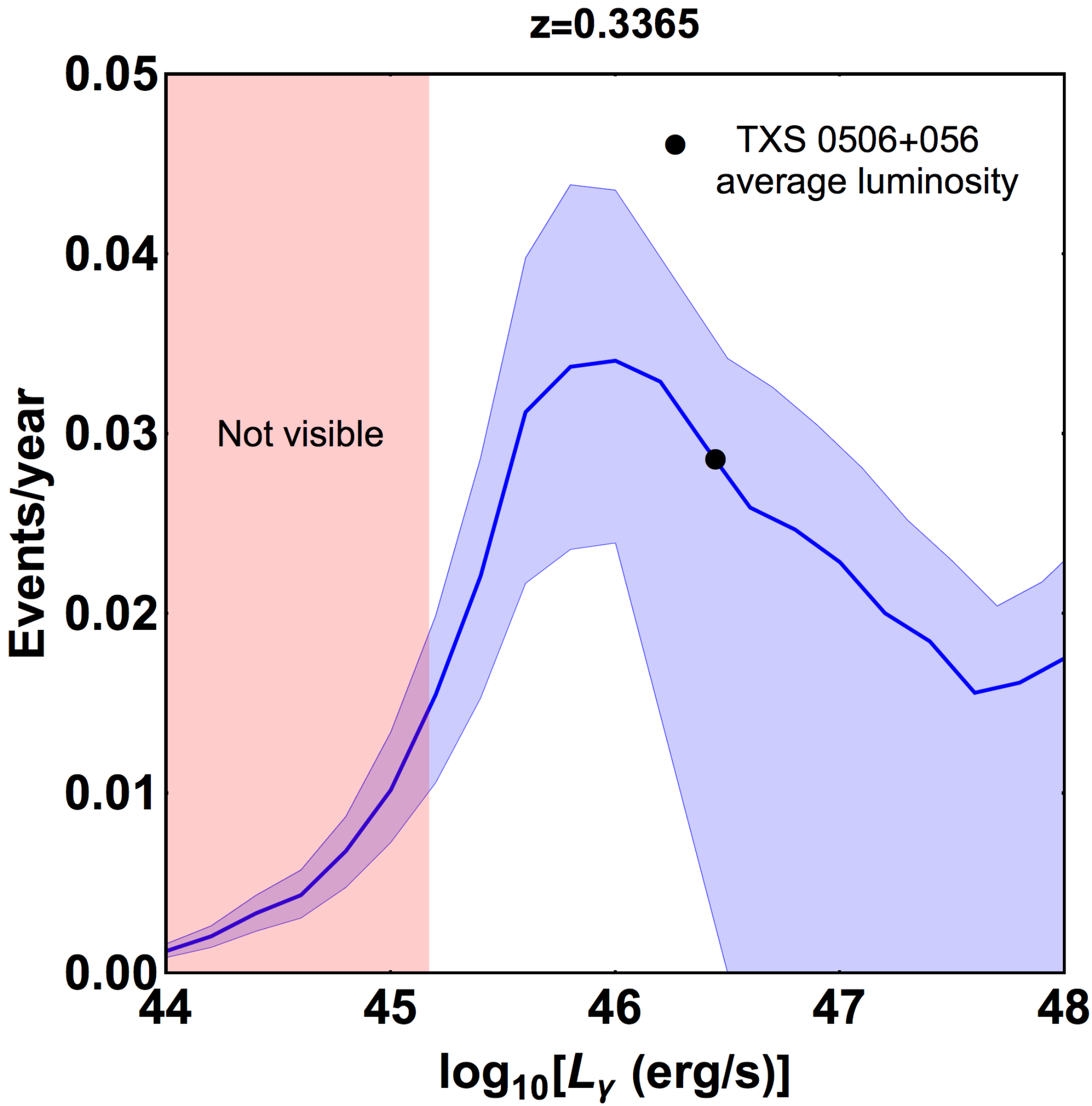}
\includegraphics[width=0.465\textwidth,angle=0]{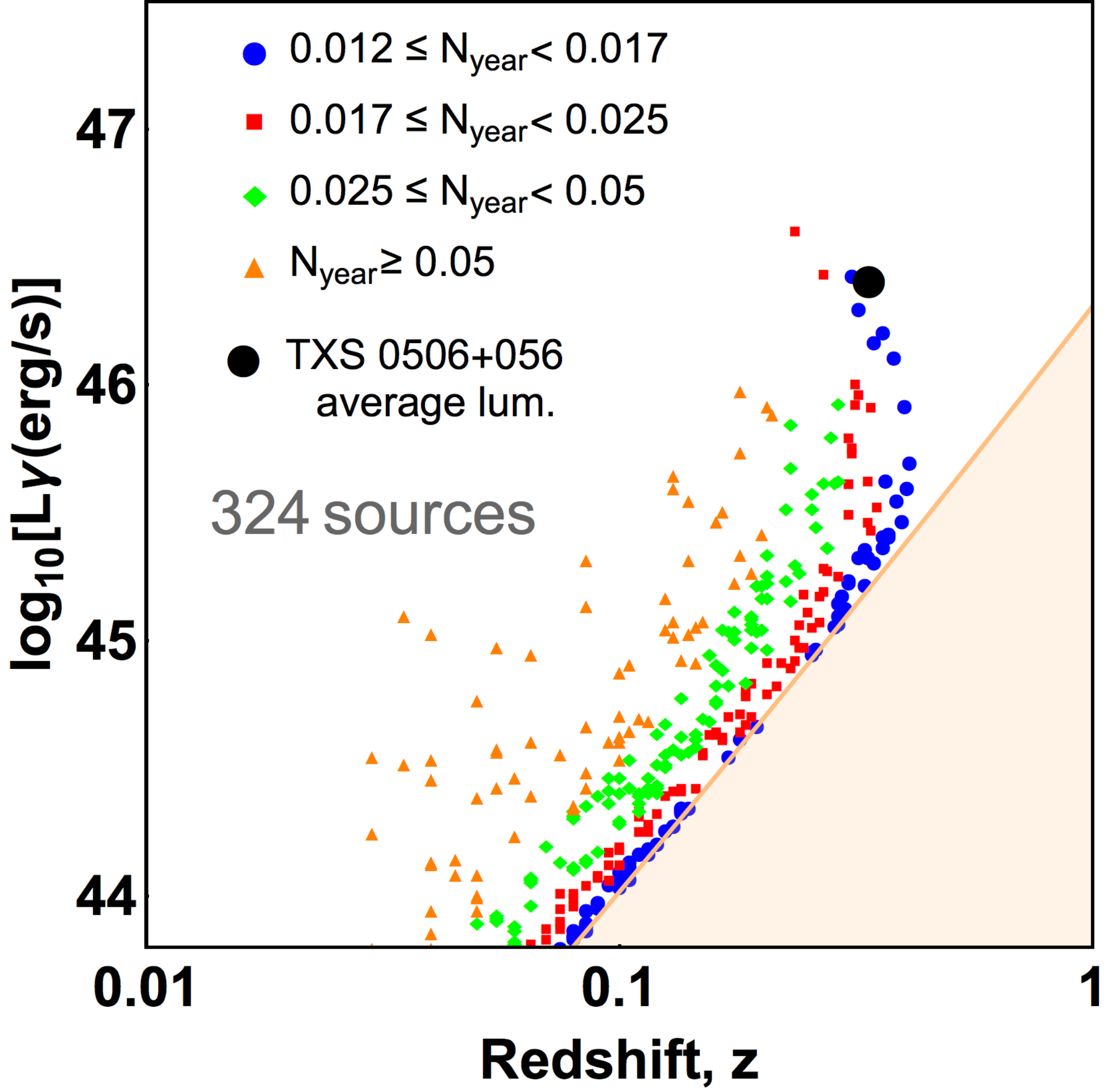} \\
\caption{Left: expected number of events produced by a single source at the same redshift as TXS 0506+056 as a function of the $\gamma$-ray luminosity (event rate refers to point source analysis). The shaded pink region denotes the unresolved sources at redshift $z=0.3365$. For a TXS-like source, the (time-averaged luminosity) number is equal to 0.012 events per year. Right: distribution of resolved sources that can produce at least the same number of events per year as TXS 0506+056. The different symbols indicate ranges of the expected number of events per year from the source.}
\label{plottxs}
\end{figure*}

\subsection{Comparison with TXS 0506+056}
\label{sec:txs}

We here discuss the implications of the neutrino associated with TXS 0506+056 in the context of our model.

\paragraph{Expected number of events from TXS 0506+056:} 
The luminosity of neutrinos is expected to be about 1\% of the $\gamma$-ray luminosity, see  middle panel of \Fig~\ref{fig:eddington}. This  is in agreement with the expectations reported in \citet{Gao:2018mnu}, in which the BL Lac has been modeled using a lepto-hadronic hybrid model. Moreover, we can evaluate the neutrino flux expected from this source, following the same procedure used in \Sec~\ref{sec:neutrinoflux} to evaluate the cumulative neutrino flux expected from the entire blazar population. We give three different expected number of events, considering:
\begin{itemize}
\item the point source analysis with events above 100 GeV. This dataset is dominated by atmospheric background \cite{icemuon};
\item the throughgoing muon sample with events above 200 TeV. This is dominated by astrophysical signal with a $\sim 30\%$ of events coming from atmospheric prompt neutrinos \cite{icemuon};
\item the alert system with Extreme High Energy Events (EHE). We will use it to compute the expected number of correlations between neutrinos and known $\gamma$-ray sources \cite{Aartsen:2016lmt}.
\end{itemize} 

Using the effective area for point sources reported in \citet{icemuon}, we obtain a (time-averaged) expected number of 0.028 events per year from TXS. This is the expected number of events considering the full energy range of the point source analysis, \ie, with deposited energy larger than 100 GeV.  In the left panel of \Fig~\ref{plottxs} we show the expected number of events per year for objects with the same redshift as TXS but different luminosities. From this plot it is clear that objects with $10^{45} \lesssim L_\gamma (\text{erg/s})  \lesssim 10^{47}$ are the best candidates to produce high-energy neutrinos, due to the fact that for lower luminosities the efficiency $\epsilon(L_\gamma)$ is lower (see \Fig~\ref{fig:efficiency}) whereas for higher luminosities the baryonic loading $\xi(L_\gamma)$ is too low (see \Fig~\ref{fig:scenario3}) and the neutrino flux is suppressed. To know how many events are expected in the throughgoing muon dataset we must use a low-energy threshold, since only muons above 200 TeV are considered there. A rough estimation can be obtained using $E_\nu=10 E_\mu$ (see the response function given in Fig. S4-S5 of~\citet{IceCube:2018cha}). We therefore define a threshold at 2 PeV for neutrinos, obtaining 0.004 events per year expected in the throughgoing muon dataset.

Let us remark that the observation of one event clearly associated to the TXS 0506+056 does not imply that the source has to emit 1 event/year. On the contrary, it is possible that there are many faint sources that give a contribution and the detection of one of them is a lucky coincidence. Indeed, the probability to detect at least one event is given by $P(>0)=1-\exp(-\mu)$, where $\mu$ is the expected rate. If $\mu \ll 1$ the previous expression becomes $P(>0) \simeq \mu$. In the assumption that $k$ sources contribute to the flux, it is sufficient that the rate of a single source is roughly $\mu \simeq k^{-1}$ in order to have a non-negligible probability to observe one event from one source. This concept is known as Eddington bias and it is explained in~\citet{Strotjohann:2018ufz} assuming different sources.

\paragraph{Expected number of total events per year:} In the right panel of \Fig~\ref{plottxs}, we show the distribution of sources that produce a number of events equal or grater compared to the TXS source.
Using the BL Lac distribution described in \Sec~\ref{sec:sourcedist} we find that there are 324 visible objects capable of producing at least as many neutrinos per year as TXS. Using the point source effective area of \cite{IceCube:2018cha} and integrating over redshift and luminosity, we find that the average number of events produced by these sources is 0.11 events per year in the point source analysis (above 100 GeV) and 0.016 events per year in the throughgoing muon dataset (above 200 TeV) for each source, which is roughly four times higher than that expected from TXS. Therefore, future correlations will be rather expected at somewhat lower luminosities and redshifts.

In order to  estimate  how many events per year we expect in IceCube from sources, we have to take into account that:
\begin{itemize}
\item there are 324 sources capable to produce at least as many neutrinos per year as TXS; 
\item assuming that these 324 sources are isotropically distributed, IceCube can only detect the ones that are visible from the Northern hemisphere, via throughgoing muons. In particular, the alert system is active for neutrinos above 500 TeV \citep{Aartsen:2016lmt} and at this energy only neutrinos coming from declination $0^\circ \leq \delta \leq 30^\circ$ can cross the Earth. Therefore, IceCube is roughly sensitive to 1/4 of the sky or, equivalently, to 1/4 of the sources.
\end{itemize}
Following the previous considerations, the expected number of signal events per year above 100 GeV $N_c^{\text{100 GeV}}$ in IceCube  is roughly
\begin{equation}
N_c^{\text{100 GeV}} \sim \frac{\rm 0.11~{\rm events}}{\rm year \times source }\times \frac{324~\text{sources}}{4} = \frac{\text{8.9 events}}{\text{year}}
\label{eq:correl}
\end{equation}
These events are hardly distinguishable from the atmospheric background, therefore it is more interesting to evaluate how many events are expected into the throughgoing muon dataset, i.e.\ above 200 TeV. It is roughly given by:
\begin{equation}
N_c^{\text{200 TeV}} \sim \frac{\rm 0.016~{\rm events}}{\rm year \times source }\times \frac{324~\text{sources}}{4} = \frac{1.3~{\rm events}}{\text{year}}
\label{eq:correl200tev}
\end{equation}
 Let us notice that this number is consistent with the hypothesis that half of the throughgoing muon flux is produced by resolved blazars, since in 8 years the contribution from these resolved objects would be 10.4 events. 
 At the present, after 8 years of exposure, 36 throughgoing muons have been observed \cite{Aartsen:2017mau}; 2/3 of them are expected to be signal events, therefore we expect roughly 12 events from resolved sources.

\paragraph{Expected number of correlations per year:} In order to compute the expected number of correlations per year, i.e.\ neutrinos associated with known objects, we need to refer to the alert system and to use the alert effective area. Using the alert effective area we obtain 0.004 events/year, on average, from each source (including redshift and luminosity distributions). Let us assume a duty cycle of 10\%, i.e.\ the sources are in flaring state only 10\% of the time, such as for TXS 0506+056 \citet{Murase:2018iyl}). Assuming no neutrino production during the quiescent state, the expected number of events would be 10 times higher during the flaring state, in order to be in agreement with the average result of our work. Therefore, assuming 0.04 events/year during the flaring state and assuming that only 10\% of the sources are in flaring state, the expected number of association is obtained in the following way:
\begin{equation}
N_c^{\text{coinc.}} \sim \frac{\rm 0.04~{\rm events}}{\rm year \times source }\times  \frac{324~{\rm{sources}}}{4} \times  0.1 = \frac{\text{0.32 events}}{\text{year}}
\label{eq:coincevents}
\end{equation}

The alert system is active since about two years \citep{Aartsen:2016lmt}, and therefore the expected number of correlations is equal to 0.64, following our baseline model. Up to now, only one correlation has been observed, which is perfectly consistent with our expectation within the Poissonian uncertainty.

\subsection{Comparison with PKS 0502+049}
\label{sec:pks}

For the sake of completeness, we tested the object PKS 0502+049 in the context of our model. In \citet{He:2018snd} the possibility that this object has emitted high-energy neutrinos, during the flaring state observed between 2014 and 2015, is discussed. See also \citet{Padovani:2018acg} for a direct comparison between this object and TXS 0506+056 as neutrino emitters. This source is an FSRQ at redshift $z=0.954$ and \citet{He:2018snd} state that to reproduce the $\gamma$-ray flare and the neutrino flare the object must exceed the Eddington luminosity by two orders of magnitude during the flaring state. Under this assumption the flaring luminosity would be $L_\gamma=9.5 \times 10^{48} \mbox{erg/sec}$. Using the previous luminosity and redshift and under scenario 3, we find that for such an object the neutrino luminosity would be $L_\nu = 7 \times 10^{-5} L_\gamma$ and it would produce a neutrino signal corresponding to the $\sim 40\%$ of the number of events per year expected from TXS 0506+056. Therefore, the association between the neutrinos observed by IceCube in 2014-2015 and the PKS 0502+049 is less plausible than the association between the same neutrinos and the TXS 0506+056 but it cannot be ruled out based on our model. 

\section{Summary and conclusion}
\label{sec:conclusion}

In this work we have studied the possibility that the diffuse astrophysical neutrino flux at the highest energies is powered by blazars, which are Active Galactic Nuclei viewed in the direction of the jet. A major obstacle has been the AGN stacking limit, which constrains the blazar contribution from the non-observation of correlations between high-energy neutrinos and known (observed) blazars, while unresolved sources may at the same time power most of the diffuse neutrino flux. Using a phenomenological relationship between spectral energy distribution (SED) of blazars and $\gamma$-ray luminosity, known as {\em blazar sequence}, and a realistic neutrino production model based on these SEDs, we have derived the implications for blazars assuming that diffuse flux and stacking limit have to be described at the same time.


We have demonstrated that the choice of a constant baryonic loading over the blazar sequence does not allow for a description of the neutrino data because, fixing the baryonic loading, high-luminosity objects are very efficient neutrino producers; as a consequence, resolved sources will dominate the neutrino flux, and the stacking limit will constrain the blazar contribution to the diffuse flux.  In order to avoid that, we have allowed the baryonic loading to change as a function of luminosity. We have analyzed two different possibilities: in the first one the ratio between luminosity in neutrinos and luminosity in $\gamma$-rays is constant (implying that the baryonic loading decreases with luminosity in a not continuous way), in the second one the baryonic loading scales continuously with the $\gamma$-ray luminosity and also the expectation from TXS 0506+056 are taken into account.
We have found that \textcolor{black}{the only scenario in which blazars can explain the high-energy neutrino flux is that where} the baryonic loading of low-luminosity objects is higher than $10^5$, whereas the baryonic loading of high-luminosity sources (both BL Lacs and FSRQs) has to be lower than $\sim$ 100. It is also possible that high-luminosity BL Lacs and FSRQs do not contribute to the neutrino flux at all,
which means that these sources may not be cosmic ray accelerators. \textcolor{black}{Under this hypothesis, low-luminosity objects} can then power the entire neutrino flux above a few hundreds of TeV, while the contribution of high-luminosity objects is limited by the stacking limit. In order to test the plausibility of our results, we have demonstrated that the baryonic loading obtained roughly satisfies the Eddington limit and the constraint coming from the non-observation of multiplets. \textcolor{black}{While previous works indicate that such high baryonic loadings are difficult to be achieved in low-luminosity BL Lacs because of their relatively inefficient accretion, one may speculate that the conditions are different if the neutrinos are produced during flares with baryon injection.} 

The recent observation of neutrinos from TXS 0506+056 can be interpreted in our baseline scenario. We find that the (time-averaged) expected number of neutrino events from an object at this redshift and luminosity is about 0.004 per year in the throughgoing muons sample. Taking into account the redshift and luminosity distributions, we find 0.016 events per blazar per year on average, coming from about 300 Fermi-LAT detected objects that are at least good neutrino emitters as TXS 0506+056. This  implies that somewhat larger event rates are expected from objects with higher neutrino luminosity and lower redshifts. Using the effective area of the alert system and taking into account the sky-coverage of IceCube, this yields about 0.3 expected neutrino--blazar associations per year.
These associations are likely to come from BL Lacs with lower luminosities $L_\gamma \sim 10^{45} \, \mathrm{erg/s}$ and redshifts $z \sim 0.1$ in our model.

In conclusion, we have demonstrated that the observed astrophysical flux of throughgoing muons and the stacking limit for blazars are \textcolor{black}{well-described only if one accepts} large enough baryonic loadings for unresolved (low-luminosity) objects, while high-luminosity BL Lacs and FSRQs are disfavored as the main contributors to the diffuse flux of high-energy neutrinos. \textcolor{black}{We have not found other possibilities compatible with all the present measurements. 
}

Let us remark that the previous considerations are not necessarily true for a single object, since FSRQs are the most efficient neutrino emitters, as discussed in \citet{Rodrigues:2017fmu} and which is evident from \figu{efficiency}, meaning that the optical thickness to photohadronic interactions is highest and the protons will transfer energy into neutrinos efficiently. For a diffuse flux, however, this efficiency has to be scaled with the baryonic loading and the number of objects of a certain class. In this work we find that the low luminosity sources are far more abundant and, given our calculations, must have higher baryonic loadings than the high-luminosity sources if AGNs are to power the diffuse neutrino flux.

As a consequence, we expect signatures of hadronic processes in the SED of low-luminosity objects, such as in X-ray or TeV $\gamma$-ray data~\citep{Gao:2016uld}.
From the population study, we expect further associations similar to the one to TXS 0506+056 at a rate of about 0.3 per year. However, associations with blazars with lower redshift and luminosity than TXS will be more likely.

{\bf Acknowledgments.}  A.P. thanks F. Vissani for fruitful past discussions concerning the connection between blazars and high-energy neutrinos. This project has received funding from the European Research Council (ERC) under the European Union’s Horizon 2020 research and innovation programme (Grant No. 646623).

\bibliographystyle{aasjournal}
\bibliography{bibliography}

\clearpage

\begin{appendices}

\section{Source model}
\label{sec:source_model}

We present here some details of the source model used in the present work. While most features are similar to \citet{Rodrigues:2017fmu}, some parameter are different, which are emphasized in the following discussion. It is also worth noting that in~\citet{Rodrigues:2017fmu} the model was applied to a previous implementation of the blazar sequence~\citep{Fossati:1998zn}, based mainly on radio and X-ray observations. The new implementation of the blazar sequence used in this work~\citep{Ghisellini:2017ico}, on the other hand, was constructed based on the more recent Fermi 3LAC blazar catalog~\citep{Ackermann:2015yfk}, as discussed in~\sect{neutrino_production}. Moreover, in~\citet{Rodrigues:2017fmu} only one sequence is considered, where all low-luminosity sources are BL Lacs and all high-luminosity sources are FSRQs, while in the present work we consider two sequences, one for BL Lacs and another for FSRQs, spanning the same luminosity range.

The magnetic field strength in the jet is in this work considered to be a soft power-law of the blazar luminosity, $B'\sim L_\gamma^{1/5}$. This scaling was employed in \App~A of~\citet{Rodrigues:2017fmu} because it can yield values of $B'$ for BL Lacs in the range 0.1-1 G, and for FSRQs in the range $1~{\rm G}$ to a few~G, which are in agreement with previous results \citep{Tavecchio:2009zb,Dermer:2013cfa}. The scaling is assumed to be the same for BL Lacs and FSRQs, which yields $B'=6.3~{\rm G}$ for the brightest FSRQ discussed in this work (see \figu{neutrinos_sequence}), and $B'=0.1~{\rm G}$ for the dimmest BL Lac. In contrast, in the main text of~\citet{Rodrigues:2017fmu} a stronger scaling was assumed, $B\sim L_\gamma^{1/2}$, which yields much weaker magnetic field strengths in the jet of low-luminosity sources.

\begin{figure*}[tbp!]
  \includegraphics[width=\linewidth]{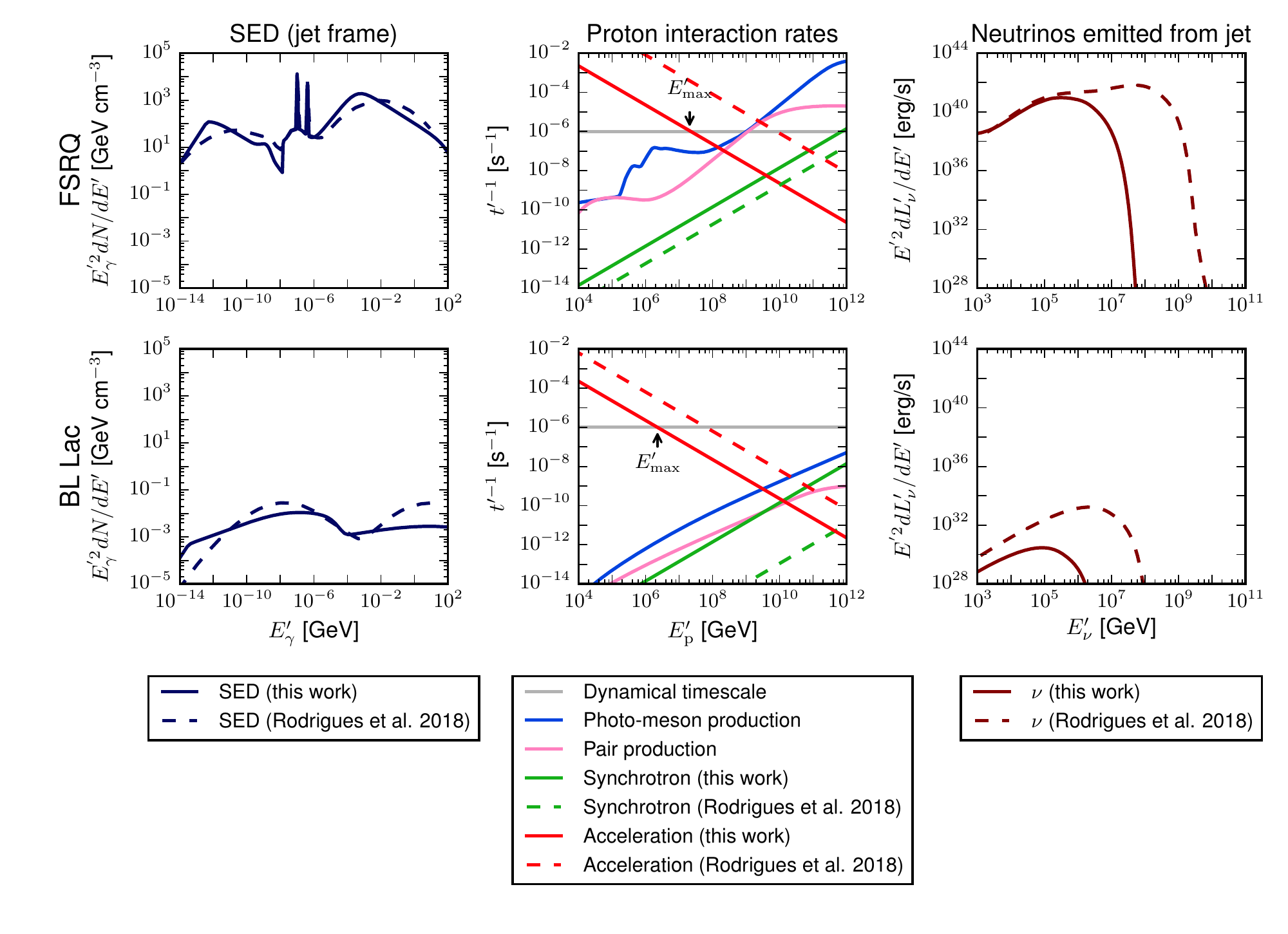}
  \caption{Left: the solid curves represent the photon density spectra in the jet blob in an FSRQ of luminosity $10^{48.5}~{\rm erg/s}$ (top) and a BL Lac with $10^{44.5}~{\rm erg/s}$ (bottom). Note how the external fields of the FSRQ appear boosted in the jet frame due to the relativistic motion of the jet (\cf~\figu{neutrinos_sequence}). For comparison, we plot in dashed the SEDs of an FSRQ and BL Lac of similar luminosities considered in~\citet{Rodrigues:2017fmu} (an FSRQ of $10^{48.8}~{\rm erg/s}$  and a BL Lac of $10^{44.6}~{\rm erg/s}$). Middle: cooling and interaction rates of protons in the jet for the same sources. The solid red line is the acceleration efficiency considered in this work, $\eta=10^{-3}$ (\equ{eta}). For comparison, the red dashed line refers to a ultra-efficient proton acceleration ($\eta=1$), discussed in~\citet{Rodrigues:2017fmu}. On the other hand, in that work lower magnetic fields were considered ($0.9~{\rm G}$ compared to $2~{\rm G}$ in this work for the FSRQ and $7~{\rm mG}$ compared to $250~{\rm mG}$ in this work for the BL Lac). This counteracts the higher acceleration efficiency in the calculation of the acceleration rates, and also reflects in lower synchrotron cooling rates (compare solid and dashed yellow lines). Right: all-flavor neutrino luminosity spectrum produced in the jet, considering the SED and acceleration efficiency used in this work (solid) and in~\citet{Rodrigues:2017fmu} (dashed). Note that in the case of the BL Lac, the suppressed neutrino production below $10^5~{\rm GeV}$ compared to~\citet{Rodrigues:2017fmu} is simply due to the higher magnetic field, which increases proton synchrotron cooling.}
\label{fig:prototypes}
\end{figure*}

In \figu{prototypes} we explore the photo-hadronic interactions in the jet of two blazars from the blazar sequence; one a high-luminosity FSRQ, and the other a BL Lac of intermediate luminosity (bottom). We compare some of their features with an FSRQ and a BL Lac of similar luminosities, but with the assumptions considered in~\citet{Rodrigues:2017fmu}. In this discussion, as well as throughout this work, we consider the case of CR escape by Bohm-like diffusion, where the rate of escape is proportional to the Larmor radius of the CRs of a given energy. As shown in~\citet{Rodrigues:2017fmu}, the assumption for the particular CR escape mechanism only affects marginally the neutrino spectra ejected by the blazar.

In the left panels of \figu{prototypes} we show the photon density spectrum in the jet. Note that in the case of the FSRQ (top panel), the external radiation is relativistically boosted into the jet blob, as discussed above ($E'_\gamma=\Gamma E_\gamma$, \cf~\figu{neutrinos_sequence}). For high-luminosity FSRQs (see end of this section), additional components of the SED are considered in addition to non-thermal radiation (assumed to be produced by electrons in the jet). These external components are an infra-red bump from a dusty torus of temperature $3000~{\rm K}$, an X-ray bump from the corona of the accretion disk~\citep{Elvis:1994}, and two broad lines emitted by molecular gas on the edge of the BLR~\citep{bl3} (\cf~\figu{model}). The broad line emission, as well as a component of the radiation from the dusty torus and the accretion disk, are assumed to isotropize in the BLR~\citep{Rodrigues:2017fmu}, and that component is relativistically boosted into the jet frame. This assumption has also been considered in previous works dealing with hadronic blazar jet models~\citep{bl3}.

In the middle panels of \figu{prototypes} we show the rates of processes undergone by protons in the jet of the FSRQ (top) and the BL Lac (bottom). The acceleration rate displayed as a solid red line is the one considered in this work, \ie,  an acceleration efficiency of $\eta=10^{-3}$ is used (\cf~\equ{eta}). In contrast, we show in dashed red the case of ultra-efficient proton acceleration, as discussed in the main text of~\citet{Rodrigues:2017fmu}. In that work, low-luminosity jets were considered to have lower magnetic field strengths, which counteracts the effect of a higher acceleration efficiency in the calculation of the acceleration rate (\cf~\equ{eta}). The maximum energy achieved by protons in the source (which is proportional to the maximum energy of the emitted neutrinos) is the energy at which acceleration becomes dominated by cooling or hadronic interactions (including adiabatic cooling, synchrotron losses, electron-positron pair production and photo-meson production). The adiabatic cooling timescale is assumed to scale with the size of the region, whose inverse timescale is represented by the gray line; note that even if the plasma blob cools slower than adiabatically, the size of the blob plays the same role as an adiabatic cooling term in that it limits cosmic-ray acceleration to the energy where the particle Larmor radius reaches the size of the region, when it cannot be efficiently accelerated further. Note also that the maximum proton energy marked in the figure corresponds to the acceleration efficiency considered in this work, while it would be 2-3 orders of magnitude higher for $\eta=1$.

As we can see by the solid curves in the right panels of \figu{prototypes}, the FSRQ (top) produces neutrinos with a maximum of $\sim10~{\rm PeV}$ in the source frame ($\sim1~{\rm PeV}$ in the jet rest frame), and the BL Lac (bottom) produces neutrinos with a maximum of 1 PeV in the source frame. If an acceleration efficiency of $\eta=1$ were considered (dashed curves), the FSRQ neutrino spectrum would instead peak at 1 EeV and the BL Lac at 50 PeV. This shows that the low acceleration efficiency considered in this work is essential to obtain a neutrino spectrum whose cutoff is compatible with the maximum energy of the observed astrophysical neutrinos. In addition, note that diffusive shock acceleration models~\citep{Inoue:2016fwn} indicate that high acceleration efficiencies may be difficult to achieve in relativistic shocks, as it is the case of blazar jets. On the other hand, with such low acceleration efficiencies, these sources cannot power the diffuse  ultra-high-energy CR spectrum (\cf~bottom panel of \figu{efficiency}). 

Besides the acceleration efficiency, the blob size is another parameter of the model that affects the peak energy and normalization of the ejected neutrino spectrum. While the value used in this work reflects a source with a one-day variability timescale, $r'_{\rm blob}=\Gamma c\times1~{\rm day}=3\times10^{16}~{\rm cm}$, other values could be considered. In low-luminosity blazars, the maximum neutrino energy then scales  $E_\nu^{\rm max}\sim r'_{\rm blob}$, because cosmic-ray acceleration is limited only by the blob size (see bottom center panel of \figu{prototypes}); in high-luminosity objects, this scaling is weaker because the maximum energy is limited by photo-hadronic interactions (see upper center panel of \figu{prototypes}). The total neutrino luminosity produced by the source scales mainly with the optical thickness to photo-meson production. The optical thickness grows linearly with $r'_{\rm blob}$ and with $n_\gamma\sim V'^{-1}\sim r'^{-3}_{\rm blob}$; thus, the normalization of the emitted neutrino spectrum scales as $r'^{-2}_{\rm blob}$.

Finally, the model geometry (briefly discussed in~\sect{neutrino_production}) follows the principles discussed in detail in~\citet{Rodrigues:2017fmu}. The jet blob is at a distance to the black hole $r_{\rm diss}=r'_{\rm blob}/\sin(\theta_{\rm jet})=\Gamma r'_{\rm blob}$ (see \figu{model}), which corresponds to the dissipation radius of the jet, assumed the same for all blazars. At the same time, the sizes of the BLR and dusty torus ($r_{\rm BLR}$ and $r_{\rm DT}$), are assumed to scale with $L_{\rm disk}^{1/2}$~\citep{bl3, Rodrigues:2017fmu}, where $L_{\rm disk}$ the accretion disk luminosity. In turn, $L_{\rm disk}$ and $L_\gamma$ also scale together as $L_\gamma\propto L_{\rm disk}^{0.683}$~\citep{bl3, Rodrigues:2017fmu}. Therefore, the jet blob lies inside the BLR only in FSRQs with luminosity $L_\gamma>2.4\times10^{48}~{\rm erg/s}$. In these cases, we model the BLR and the dusty torus as two additional radiation zones. The CRs that escape the jet then interact with the photons fields present in either zone before leaving the source. For FSRQs with $r_{\rm BLR}<r_{\rm diss}<r_{\rm DT}$, \ie, with luminosity $3\times10^{46}<L_\gamma~{\rm (erg/s)}<2.4\times10^{48}$, radiation from the dusty torus is accounted for in the calculation of CR interactions in the jet, but a one-zone model is employed, such as for BL Lacs and dimmer FSRQs (this is the case for those SEDs in \figu{neutrinos_sequence} in which the thermal IR bump, but no optical bumps or emission lines, are present). This approximation is justified by the conclusion in~\citet{Rodrigues:2017fmu} that for a purely diffusive cosmic-ray escape, the neutrinos produced during CR propagation through the BLR typically contribute negligibly to the total neutrino fluence. Rather, the flux is dominated either by the neutrinos produced directly in the jet, or during CR propagation through the thermal field from the dust torus.

\section{The distribution of BL Lacs and FSRQs}
\label{sec:ajello}

The parametrization in~\citet{Ajello:2013lka, Ajello:2011zi} describes the differential source distribution

\begin{equation}
\frac{d^3N}{dV_c \,dL_\gamma \,d\Gamma},
\label{eq:distrov}
\end{equation}
which represents the number of sources $N$ per comoving volume $V_c$, $L_\gamma$ is the emitted luminosity between 0.1 and 100 GeV, and $\Gamma$ is the slope of the $\gamma$-ray flux, assuming power law spectra in that energy interval. 

For our purpose, it is convenient to write the previous expression as a function of the redshift $z$ and $\ell \equiv \log_{10}[\frac{L_\gamma}{\rm{erg/s}}]$. The comoving volume is given by:
\begin{equation}
V_c(z)= \frac{4}{3} \pi \, D_c(z)^3 \, .
\end{equation}
The comoving distance $D_c(z)$ is defined as follows:
\begin{equation}
D_c(z)= D_H \times d(z)
\label{eq:comdist}
\end{equation}
where $H_0=4.22~\text{Gpc}$ is the Hubble distance. The function $d(z)$ is given by
\begin{equation}
d(z)=\int_0^z \frac{dz'}{h(z')} \, ,
\end{equation}
where $h(z)=\sqrt{\Omega_\Lambda+\Omega_m(1+z)^3}$, with $\Omega_\Lambda=0.73$, $\Omega_m=0.27$. Based on the previous equations, the relation between the comoving volume and the redshift is
\begin{equation}
J_z(z) \equiv \frac{dV_c}{dz} = 4\pi \, D_H^3 \, \frac{d^2(z)}{h(z)} \, .
\end{equation}
The Jacobian obtained from the transformation $L_\gamma \rightarrow \ell$ is given by:
\begin{equation}
J_\ell(\ell) \equiv \frac{dL}{d\ell}=\ln(10) \ L_\gamma \, .
\end{equation}
Combining the two previous expressions with \equ{distrov} we obtain:
\begin{equation}
\frac{dN}{dz d\ell d\Gamma} = J_z(z) \times J_\ell(\ell) \times \frac{dN}{dV_c dL_\gamma d\Gamma} \, .
\label{eq:conversion}
\end{equation}
\begin{table*}[t]
\caption{Table of the parameters that define the distribution of BL Lacs and FSRQs. Only the first two parameters are dimensional, as indicated.}
\begin{center}
\begin{tabular}{ccccccccccccc}
\hline
& A [$\text{Gpc}^{-3}$] & $L_* [\text{erg/s}]$ & $\gamma_1$ & $\gamma_2$ & $\mu_*$ & $\beta$ & $\sigma$ & $p_1^*$ & $p_2^*$ & $\tau$ & $z_c^*$ & $\alpha$ \\
\hline
BL Lacs & 3.39 & $10^{47.4472}$ & 0.27 &1.86 & 2.10 & 0.0646 & 0.26 & 2.24 & -7.37 & 4.92 & 1.34 & 0.0453  \\
FSRQs & 3.06 & $10^{47.9243}$ & 0.21&1.58 & 2.44 &0 &0.18 &7.35 &-6.51 &0 &1.47 &0.21  \\
\hline
\end{tabular}
\end{center}
\label{tab:distro}
\end{table*}%
The parametrization by Ajello et al. is characterized by several parameters, which we report in \tabl{distro}. Only the first two parameters, namely $A$ and $L_*$, are dimensional, respectively in $\rm{Gpc}^{-3}$ and erg/s. In this discussion, the quantity $L_\gamma$ is always in units of $\text{erg/s}$.

The distribution in the slope $\Gamma$ is assumed to be Gaussian, 
\begin{equation}
g(\Gamma,L_\gamma)=\exp \left[ -\frac{(\Gamma-\mu(L_\gamma))^2}{2\sigma^2} \right] \, ,
\end{equation}
where $\mu(L_\gamma)=\mu_* + \beta [\log_{10}(L_\gamma)-46]$ \, .

The distribution in luminosity is given by a double power law:
\begin{equation}
f(L_\gamma)=\frac{1}{\left(\frac{L_\gamma}{L_*}\right)^{\gamma_1}+\left(\frac{L_\gamma}{L_*}\right)^{\gamma_2}} \, .
\end{equation}

The distribution in $z$ is described by an evolution factor, also a double power law:
\begin{equation}
e(\omega,L_\gamma)= \frac{1}{\omega^{-p_1(L_\gamma)}+\omega^{-p_2}} \ \ \ \mbox{with } \omega(z,L_\gamma)=\frac{1+z}{1+z_c(L_\gamma)}\, ,
\label{eqwrong}
\end{equation}
where $z_c(L_\gamma)=z^*_c \times (L_\gamma/10^{48})^\alpha$ and $p_1(L_\gamma)=p_1^* + \tau[\log_{10}(L_\gamma-46)]$.
\textbf{Note: in \citet{Ajello:2011zi,Ajello:2013lka} the previous equation is reported with a wrong positive sign for both $p_1$ and $p_2$. This mistake has been corrected later, in the footnote of \citet{Ajello:2015mfa}. The wrong sign produces important difference especially in the distribution of low luminosity BL Lacs, that accumulates at high redshift using the wrong distribution.}

Summarizing, we have: 
\begin{equation}
\frac{d^3 N}{dV_c dL_\gamma d\Gamma} = \frac{A}{\ln(10) L_\gamma} \ g(\Gamma,\ell) f(\ell) e(\omega, \ell) \, ,
\end{equation}
which, using \equ{conversion}, becomes:
\begin{equation}
\frac{d^3 N}{dz \, d\ell \, d\Gamma} = \mathcal{N} \times g(\Gamma,\ell) \, f(\ell) \, e(\omega, \ell) \times \frac{d(z)^2}{h(z)} \, .
\end{equation}
The coefficient $\mathcal{N}=4\pi D_H^3 \times A$ is a dimensionless number.

Our purpose is to compute the diffuse high-energy neutrino flux from blazars. Since the neutrino spectra at the source only depends on the luminosity, we can integrate over the slope $\Gamma$ (we cannot integrate over the redshift $z$, since the redshift enters in the computation of the neutrino flux at Earth). Henceforth we will use the distribution of BL Lacs and FSRQs defined as follows:
\begin{equation}
\frac{d^2 N}{dz \, d\ell} = \int_{\Gamma_1}^{\Gamma_2} \frac{d^3 N}{dz\, d\ell \, d\Gamma} d\Gamma \, .
\label{eq:dndzdl}
\end{equation}
The intervals of integration for BL Lacs are $z \in [0.03,6]$ and $\ell \in [43.85,52]$, and for FSRQs $z \in [0,6]$ and $\ell \in [44,52]$. The distributions $\frac{d^2N}{dz d\ell}$ of BL Lacs and FSRQs are represented in the left panel of \figu{distro}.

\end{appendices}

\end{document}